\documentclass[12pt]{article}
\usepackage{amssymb,amsmath,amscd}
\usepackage{epsfig}
\usepackage{epstopdf}
\usepackage{latexsym}
\usepackage{graphicx}
\usepackage{subfigure}
\usepackage{booktabs}
\usepackage{bbm}
\usepackage[margin=20pt,small]{caption}
\usepackage[vcentermath,stdtext]{youngtab}
\usepackage[toc]{appendix}
\usepackage[numbers,compress]{natbib}

\usepackage{color}
\usepackage{datetime}
\usepackage[
      colorlinks=false,
      linkcolor=darkblue,  
      urlcolor=blue,    
      filecolor=blue,     
      citecolor=red,
linktocpage=true,
      pdfstartview=FitV,
      bookmarksopen=true    
      ]{hyperref}

\ifpdf
\fi

\def\coeff#1#2{\relax{\textstyle {#1 \over #2}}\displaystyle}

\def\cA{{\cal A}}
\def\cB{{\cal B}}

\def\cM{{\cal M}}
\def\cN{{\cal N}}

\def\cP{{\cal P}}

\def\cV{{\cal V}}

\def\eql{=}

\def\RR{\mathbb{R}}

\definecolor{cardinal}{rgb}{0.6,0,0}
\definecolor{darkgreen}{rgb}{0,0.5,0}
\definecolor{golden}{rgb}{0.92, 0.7, 0}
\definecolor{midnight}{rgb}{0, 0, 0.5}
\definecolor{darkblue}{rgb}{0.2, 0, 0.8}

\def\bfs#1{{\boldsymbol #1}}

\def\cals#1{\mathcal{#1}}
\def\cA{{\cals A}}

\def\eo{\overset{_{\phantom{.}\circ}}{e}{}}
\def\Go{\overset{_{\phantom{.}\circ}}{\Gamma}{}}
\def\go{\overset{_{\phantom{.}\circ}}{g}{}}
\def\Do{\overset{_{\phantom{.}\circ}}{D}{}}

\def\cBo{\overset{_{\phantom{.}\circ}}{\cals B}{}}
\def\cAo{\overset{_{\phantom{...}\circ}}{\cals A}{}}
\def\be{\bar e}

\def\bxi{{\bfs \xi}}

\def\Ga{\Gamma}

\begin{document}  

\begin{titlepage}
\flushright{AEI-2011-093} 
\bigskip
\bigskip

\bigskip
\bigskip
\begin{center} 
{\Large \bf  Consistent truncation of $\bfs d\,\bfs =\,\bfs 1\bfs 1$ supergravity on ${\bfs A\bfs d\bfs S_{\bfs 4}} \times  \bfs S^{\bfs 7}$}

\bigskip\bigskip\bigskip\bigskip

{\bf Hermann Nicolai${}^{(1)}$  and
 Krzysztof Pilch${}^{(2)}$  \\ }
\bigskip
${}^{(1)}$
Max-Planck-Institut f\"ur Gravitationphysik\\
Albert-Einstein-Institut\\
M\"uhlenberg 1,
D-14476 Potsdam, Germany
\vskip 5mm
${}^{(2)}$ Department of Physics and Astronomy \\
University of Southern California \\
Los Angeles, CA 90089, USA  \\
\bigskip
Hermann.Nicolai@aei.mpg.de, pilch@usc.edu  \\
\end{center}

\vspace{1cm}
\begin{abstract}
{
\small
We study the system of equations derived twenty five years ago by B.~de Wit and  the first 
author [Nucl.~Phys.~{B281} (1987) 211] as  conditions for the consistent truncation of 
eleven-dimensional supergravity on $AdS_4\times S^7$ to gauged $\cN\!=\!8$ supergravity 
in four dimensions. By exploiting the $\rm E_{7(7)}$ symmetry, we determine the most general 
solution to this system at each point on the coset space $\rm E_{7(7)}/SU(8)$. We show that  
invariants of the general solution are given by the fluxes in eleven-dimensional 
supergravity. This allows us to both clarify the explicit non-linear ans\"atze for the 
fluxes given previously and to fill a gap in the original proof of the consistent truncation. 
These results are illustrated with several examples.
}
\end{abstract}

\end{titlepage}


\tableofcontents

\section{Introduction}

Among the known examples of consistent non-linear embeddings in Kaluza-Klein supergravity
(and Kaluza-Klein theories in general), the non-linear embedding of $\cN\!=\!8,\, d\!=\!4$ 
gauged supergravity \cite{de Wit:1982ig} into $\cN\!=\!1, d\!=\!11$ supergravity 
 \cite{Cremmer:1978km} stands out as the most subtle and complicated. This embedding 
was  derived a long time ago  in \cite{deWit:1986iy} on the basis of the SU(8) invariant reformulation of $d\!=\!11$ supergravity presented in \cite{deWit:1986mz} (a list
of references to earlier work can be found in \cite{deWit:1986iy}).
However,  with the exception of \cite{hep-th/9911238}, 
where the simpler embedding of maximal $d\!=\!7$ gauged supergravity into the 
$d\!=\!11$ theory was completely worked out,  there has not been much 
follow-up work on maximal supergravity embeddings since then. In particular, 
no complete proof  exists for  the $AdS_5 \times S^5$ compactification of IIB supergravity
to maximal gauged $\cN\!=\!8$ supergravity in $d\!=\!5$,   although partial
formulae for the embedding were obtained in \cite{hep-th/9812035,hep-th/0006066,hep-th/0106032}. By contrast, there has been considerable work on consistent truncations of $\cN\!=\!1$, $d\!=\!11$ supergravity to  non-maximal supergravities in $d\!=\!4$, whose 
scalar sectors are much simpler.\footnote{For  a partial 
 list of references, see  \cite{hep-th/0003286,hep-th/9910252,hep-th/0002099,arXiv:0712.3560,arXiv:0901.0676,arXiv:1008.1423,arXiv:1105.6114,arXiv:1106.4781,arXiv:1110.5327}.
}

In this paper we re-analyze 
the  embedding of the scalar sector of the $\cN\!=\!8$ theory and, in particular, examine the flux ans\"atze in \cite{deWit:1986iy} for solutions of $\cN\!=\!1$,  $d\!=\!11$ supergravity that correspond to lifts of critical points of the scalar potential in four dimensions. 
Our present interest in this problem  has been motivated on the one hand by the recent 
discovery  of a large number of new critical points \cite{Fischbacher:2009cj,Fischbacher:2010ec,Fischbacher:2011jx,DallAgata:2011aa} for which the corresponding eleven-dimensional solutions are not yet known. On the other hand,  the explicit flux formulae in \cite{deWit:1986iy} have never 
been tested for any but the maximally supersymmetric point. The present investigation
originated from an attempt to extend this analysis to non-trivial vacua, and to test the
general formulae by performing numerical checks for some configurations of the 
scalar fields. To our surprise, these checks revealed  systematic 
inconsistencies.\footnote{Recently, a similar inconsistency for the $\rm SO(7)^+$ 
 point was independently observed in \cite{Ahn:2011nh}.} 
This raised questions not just about the flux formulae {\it per se},  but also about the 
completeness of the proof of consistency for the $S^7$ truncation in \cite{deWit:1986iy}.  
In this paper, we resolve the apparent discrepancies and complete the proof of consistency,
on the way also deriving the exact formulae for the non-linear flux ans\"atze. Indeed, 
the formulae given in \cite{deWit:1986iy}, after considerable work, turn out to be 
essentially correct, {\em modulo} an important subtlety that was not appreciated 
there, and which is the main subject of the present paper.

Recall that solutions of $d\!=\!11$ supergravity corresponding to given critical points 
of gauged $\cN\!=\!8$ supergravity are warped products   $AdS_4\times \cM_7$, 
\begin{equation}\label{themetr}
ds_{11}\eql \Delta^{-1}ds_{AdS_4}^2+\,ds_{\cM_7}^2\,,
\end{equation}
\begin{equation}\label{fluxsol}
F_{(4)}\eql  f\,\Delta^{-2}\,{\rm vol}_{AdS_4}+\coeff 1 {4!}\, F_{abcd}\, e^a\wedge e^b\wedge e^c\wedge e^d\,,
\end{equation}
where $ds^2_{AdS_4}$ denotes the line element in $AdS_4$ with warp factor 
$\Delta^{-1}$, and $ds_{\cM_7}^2 \equiv  g_{mn} \,dy^m \otimes dy^n$ is the internal 
seven-metric (as in \cite{deWit:1986iy}, we will label four-dimensional coordinates $x^\mu$ 
by Greek indices $\mu,\nu...= 0,1,2,3$, and internal coordinates $y^m$ by Latin indices 
$m,n,...=1,...,7$).  The flux components are defined in the usual way, with {\em flat} indices
and $24i f \equiv \varepsilon^{\alpha\beta\gamma\delta} F_{\alpha\beta\gamma\delta}$
(so the constant $f_0\equiv f\Delta^{-2}$ is the Freund-Rubin parameter \cite{EFI 80/35-CHICAGO}). Hence, the internal manifold $\cM_7$ is a deformation of the seven-sphere 
$S^7$ (which corresponds to the maximally supersymmetric vacuum \cite{DP}). 
In fact, such deformations can be studied for {\em any} field configuration of the $d\!=\!4$ theory satisfying the field equations of $\cN\!=\!8$ supergravity, in which case the internal metric and fluxes depend on both $x$ and $y$, such as for instance the $AdS_4$-type vacua with
$x$-dependent scalar field configurations which have  attracted recent interest 
in the context of M2-branes and holographic superconductors. The main question 
then concerns $(i)$ {\em how} to construct the non-linear embedding of a given $d\!=\!4$ configuration into the  $d\!=\!11$ theory, and $(ii)$ the consistency of this embedding. 
By a {\em consistent truncation} (or embedding) we shall here generally 
mean that any $d\!=\!4$ solution (whether $x$-independent or not), 
when embedded into the $d\!=\!11$ theory, should yield an exact solution 
of the latter {\em at the full non-linear level} (see e.g. \cite{Gibbons} for an introductory review). 
As we will see, this requirement will lead to rather complicated formulae for both 
the internal metric and the fluxes in terms of the $d\!=\!4$ fields. 

\section{Synopsis}
\setcounter{equation}{0}
In this section we recall  some central results from earlier work and summarize our 
main new insights. We strongly recommend that readers consult the two main 
references \cite{deWit:1986iy,deWit:1986mz}, whose notations and conventions 
we will follow throughout this paper, as well as \cite{deWit:1984nz} 
for further details whenever necessary.

The  explicit construction of the lift in  \cite{deWit:1984nz,deWit:1986mz,deWit:1986iy} 
starts with the so-called {\em generalized vielbein}.  
This object is  a `soldering form' with one internal upper world index and two flat
(tangent space) $\rm SU(8)$ indices, and plays a key role in the $\rm SU(8)$ invariant
reformulation of $d\!=\!11$ supergravity presented in \cite{deWit:1985iy,deWit:1986mz}.
It  is expressed in two different and independent ways,  one coming from the $d\!=\!11$ side 
via the reformulation \cite{deWit:1986mz}, and the other coming from the $d\!=\!4$ theory, 
and  in terms of the scalar fields of $\cN=8$ supergravity and the  $S^7$ Killing spinors. 
The comparison between the $d\!=\!4$ and the $d\!=\!11$ expressions, obtained 
by judicious analysis of the supersymmetry variations, then yields crucial information about
the non-linear embedding, as we now explain.

Let us start with the $d\!=\!11$ side, which is based on the reformulation \cite{deWit:1986mz},
where the original tangent space symmetry $\rm SO(1,10)$ is  replaced by 
$\rm SO(1,3) \times SU(8)$, as appropriate to a (4+7)-decomposition of the original 
theory~\cite{Cremmer:1978km}, and where the dependence on all coordinates is initially
retained. 
The generalized vielbein is defined from the $d\!=\!11$ supersymmetry variations as
\begin{equation}\label{genvield11}
e^m_{AB}(x,y) \eql i\,e_a{}^m\,\Delta^{-1/2}\,  \Gamma^a _{AB}\,,
\end{equation}
where $e_m{}^a(x,y)$ is the siebenbein of the full metric on $\cM_7$, and
$e_a{}^m(x,y)$ its inverse. The factor $\Delta$ is essentially the siebenbein 
determinant, except that for convenience we define
\begin{equation}\label{S}
S_a{}^b(x,y) ~\equiv~ \eo_a{}^m(y) e_m{}^b(x,y) \;,\qquad \Delta~\equiv~\det S\,,
\end{equation}
thus taking out the $y$-dependent background factor $\det \eo_m{}^a$,  where 
$\eo_m{}^a(y)$ is the background $S^7$ siebenbein, and the metric on the round 
$S^7$ is $\go_{mn}\eql \eo_m{}^a \eo_n{}^b\delta_{ab}$.\footnote{Factoring out the
  background $S^7$ siebenbein leads to extra determinant factors $\go$ in various
  formulae below. Such factors can be dropped for all practical purposes by adopting a 
  local frame where $\eo_m{}^a = \delta_m{}^a$.}  The $\rm SO(7)$ gamma 
matrices, $\Gamma^a$, are purely imaginary, and therefore $e^m_{AB}$, as defined in 
\eqref{genvield11}, is real. However, a crucial step taken in \cite{deWit:1986mz} 
in order to re-write the theory into SU(8) covariant form is now to replace \eqref{genvield11} 
by the more general definition
\begin{equation}\label{genvield11a}
e^m_{AB}(x,y)  \eql  i\,e_a{}^m\,\Delta^{-1/2}\,  (\Phi^T\Gamma^a\Phi) _{AB}
\; , \quad e^{mAB} = (e^m_{AB})^*\,,
\end{equation}
where $\Phi^A{}_B(x,y)$ is an arbitrary {\em local} SU(8) rotation depending on all
eleven coordinates. In this way the local 
SO(7) tangent space symmetry is enhanced to local SU(8) in eleven dimensions. As a consequence, the {\em real\/} internal siebenbein is converted into a {\em complex\/} object transforming under local $\rm SU(8)$, unlike the original siebenbein which transforms 
only under $\rm SO(7)$. The real form \eqref{genvield11} is then viewed as an 
SU(8) tensor ${\bf 28}\oplus \overline{\bf 28}$ taken in a special gauge.\footnote{In fact,
  the complex pair $(e^m_{AB}, e^{mAB})$ can be assigned to the $\bf{56}$ representation
  of E$_{7(7)}$, even though the latter is only a symmetry of the theory when compactified
  on $\mathbb{R}^{1,3}\times T^7$ \cite{LPTENS 79/6}.} This gauge choice will 
prove extremely useful below, and we will return to it on several occasions.

On the $d\!=\!4$ side, by contrast, the generalized vielbein is determined in terms of the 
70 scalar fields of $\cN\!=\!8$ supergravity  and the 28 Killing vectors on  $S^7$  as
\begin{equation}\label{genvieldef}
e^m_{\,ij}(x,y) \eql 
K^{m\,IJ}(y)
\,\big(u_{ij}{}^{IJ}(x)+v_{ijIJ}(x)\big)\,,\qquad e^{m\,ij}\eql(e^m_{\,ij})^*\,,
\end{equation}
where the scalar `56-bein'
\begin{equation}\label{V}
\cals V (x)\eql \left(\begin{matrix}
u_{ij}{}^{IJ} (x)& v_{ijIJ}(x)\\ v^{ijIJ}(x) & u^{ij}{}_{IJ}(x)
\end{matrix}\right)\in {\rm E_{7(7)}}\,,
\end{equation}
is an element of the maximally (`split') non-compact form of the $\rm E_7$ Lie group in the 
fundamental representation. The  Killing vectors  in \eqref{genvieldef} are represented 
in the usual way as bilinears of the  Killing spinors, 
\begin{equation}\label{Kvectdef}
K^{m\,IJ}\eql i\,\eo_a{}^m\,\bar\eta^I\Gamma^a\eta^J\,.
\end{equation}
Those  (commuting) Killing spinors, $\eta^I(y)$, satisfy
\begin{equation}\label{kspeqs}
(\Do_m+\coeff i 2\,m_7\,\Go_m)\,\eta^I\eql 0\,,\qquad I=1,\ldots,8\,,
\end{equation}
where $m_7$ is the inverse  radius of $S^7$, $\Go_m\equiv \eo_m{}^a\Gamma_a$,  and $\Do_m$ denotes the $S^7$ background covariant derivative. 

We also note that, when considered as $8\times 8$ matrices, the Killing spinors are
orthonormal, in the sense that
 $\eta^I{}_A \eta^A{}_J = \delta^I{}_J$, {etc.} As explained in \cite{deWit:1986iy}, this allows 
us to use the Killing spinors to convert the two kinds of $\rm SU(8)$ indices: $A,B,C,\dots$, 
and $i,j,k,\dots$ or $I,J,K,\ldots\,$,   appropriate to $d\!=\!11$ and $d\!=\!4$, respectively, into one another. 
 However,  a direct comparison between $d\!=\!11$ and $d\!=\!4$ quantities is  more subtle, and realizing that was one of the crucial steps in the proof of the consistent truncation in 
\cite{deWit:1986iy}.

One  key ingredient in the proof  of consistency is the $\rm SU(8)$ rotation matrix 
$\Phi\equiv \Phi(x,y)$ introduced in \eqref{genvield11a}, which is required  for a consistent `alignment' of the $d\!=\!4$ and $d\!=\!11$ theories and, in particular, of the generalized 
vielbeine \eqref{genvield11a} and \eqref{genvieldef}. More precisely, the $d\!=\!4$ 
and $d\!=\!11$ vielbeine \eqref{genvield11a} and \eqref{genvieldef} above are related by
\begin{equation}\label{aligngv}
e^m_{AB}(x,y)  \equiv  i\,e_a{}^m\,\Delta^{-1/2}\,  \Gamma^a_{CD}
\,\Phi^C{}_A\Phi^D{}_B\eql e^m_{\,ij}\,\eta^i{}_A\eta^j{}_B\,.
\end{equation}
This formula makes obvious the necessity of complexifying the original internal  siebenbein 
\eqref{genvield11} via \eqref{genvield11a}, because \eqref{genvieldef} and hence the right hand 
side of \eqref{aligngv} are manifestly complex. It also confirms that the SU(8) rotation 
$\Phi$ in general depends non-trivially on {\em both\/}  the $d\!=\!4$ (space-time) 
and the $d\!=\!7$ (internal) coordinates. The existence of  the SU(8) rotation $\Phi$  for 
{\em any} vielbein of the form \eqref{genvieldef} follows from the fact that the latter can 
be shown to satisfy the Clifford property by virtue of some E$_{7(7)}$ identities, as explained in 
Section~2 of  \cite{deWit:1986iy}.

From the two different representations of the generalized vielbein 
in \eqref{genvield11} and \eqref{genvieldef}, and from \eqref{aligngv},
we deduce two key results:
\begin{itemize}
\item [(i)] The {\em non-linear metric ansatz} \cite{deWit:1984nz}
\begin{equation}\label{metrivfor}
8\,(\Delta^{-1}\,g^{mn})(x,y) 
\eql e^m_{\,ij}e^{n\,ij}
= (K^{mIJ} K^{nKL})(y) (u_{ij}{}^{IJ} + v_{ijIJ})(u^{ij}{}_{KL} + v^{ijKL})(x)
\,,
\end{equation}
implicitly giving the dependence of the internal metric on the scalar 56-bein
$\cals V (x)$ and the $S^7$ Killing vectors $K^{IJ}(y)$.  Given any
 configuration of the $\cN\!=\!8$ fields, this formula can be solved (at least in 
 principle) for the embedded internal metric $g_{mn}(x,y)$.
\item [(ii)]  The $\rm SU(8)$ rotation matrix $\Phi(x,y)$ is determined as a function of 
  the scalar 56-bein and the $S^7$ Killing spinors. As already mentioned, this 
  $\rm SU(8)$ rotation is needed to `align' the linear fermionic 
  ans\"atze with the non-linear bosonic ones in an $\rm SU(8)$ gauge where the 
  linear fermionic ans\"atze are exact to all orders, as explained in \cite{deWit:1986iy}.
\end{itemize}

Of course, closed form expressions for either $g_{mn}(x,y)$ and $\Phi(x,y)$ are hard 
to come by, and can only be obtained in very special circumstances. Nevertheless, the 
non-linear metric ansatz (\ref{metrivfor}) has been successfully tested over the 
years for a variety of non-trivial solutions: critical points \cite{deWit:1984nz,Corrado:2001nv,BKPWtoappear}, 
RG flows \cite{Corrado:2001nv,hep-th/0112010,hep-th/0208137,hep-th/0212190,hep-th/0304132} and quadratic fluctuations \cite{Bobev:2010ib}. It was also used to construct smaller truncations \cite{hep-th/0002099,Bobev:2010ib}, and was generalized to maximal supergravities in $d\!=\!5$ \cite{hep-th/9812035} and $d\!=\!7$ \cite{hep-th/9911238,hep-th/0002028}. 
Observe that \eqref{metrivfor} fixes the overall normalization of the metric relative to the
trivial vacuum for {\em any} solution of the $d\!=\!11$ equations of motion corresponding 
to a consistent embedding of an on-shell configuration of $N\!=\!8$ supergravity (whereas
the former are in principle only determined up to an overall scaling).

Similarly, the $\rm SU(8)$ rotation is known in closed form only for some special critical 
points, the very simplest example being the maximally supersymmetric point, $\Phi=1$. 
Corrections to first order in the supersymmetry parameter induced by the non-linear
embedding were given in \cite{ANP}, although without mention of SU(8).
For purely scalar fluctuations, with no pseudoscalars, a  perturbative expansion  
for $\Phi$  was derived  in \cite{Nilsson:1984dj}, but no closed form for the summed
series is known. For all other scalar and pseudoscalar configurations, the explicit 
solutions for $\Phi$  become rapidly very complicated and cumbersome,  
as can be seen from the examples in  Section~\ref{secsix}.  
 
While the above results are enough to derive the non-linear metric ansatz, they are not
sufficient to obtain the fluxes as functions of the scalar 56-bein. For this we need to invoke 
the extra information provided by two consistency requirements, namely $(i)$ the 
generalized vielbein postulate, and $(ii)$ the so-called $\frak A$-equations.\footnote{Let 
  us, however, emphasize that the final formulae for the  non-linear flux ans\"atze can only 
  be valid {\em on-shell} because the dualizations needed to convert the two-form
  fields from $d\!=\!11$ supergravity to scalar fields necessarily require the equations
  of motion. This is in marked contrast to the $AdS_7 \times S^4$ truncation of
  Ref.~\cite{hep-th/9911238}, where the scalar fields arise directly in the reduction 
  without dualizations, whereas similar complications can be anticipated for the 
  $AdS_5\times S^5$ truncation which requires the dualization of a three-form field.}

The first  condition  derived in  \cite{deWit:1986mz} is that the generalized vielbein 
must satisfy  an equation, the so-called {\em Generalized Vielbein Postulate} 
(or {\em GVP}, for short). Like the generalized vielbein itself, this condition comes in two
different guises. Again, we first discuss the equation as obtained 
from the $d\!=\! 11$ side where $e^m_{AB}$ must  obey
\begin{equation}\label{vieleqs11}
\Do_me^n_{AB}+\cals B_m{}^{C}{}_{[A}e^n_{B]C}+\cals A_{m\,ABCD}e^{n\,CD}\eql 0\,,
\end{equation}
with a corresponding equation for the complex conjugate vielbein $e^{mAB}$.
Here, $\cals B_m{}^A{}_B(x,y)$ and $\cals A_{m\,ABCD}(x,y)$ together can be viewed as an 
${\rm E_{7(7)}}$ connection in the seven internal dimension. The {\it GVP\/} constitutes
the analog of the corresponding conditions for the $d\!=\!4$ connection
$(\cals B_\mu{}^A{}_B \,,\, \cals A_{\mu ABCD})$, which upon compactification on
$S^7$ reduce to the Cartan-Maurer  equations for ${\rm E_{7(7)}}$ (with $\rm SO(8)$ 
gauge covariant derivatives, cf. \cite{deWit:1986iy}). 

Explicit expressions in terms of $d\!=\!11$ fields are obtained by careful analysis 
of the $d\!=\!11$ supersymmetry variations \cite{deWit:1986mz}:
\begin{align}\label{su8ABB}
\cals B_m{}^A{}_B & \eql\coeff 1 2 \,(S^{-1} \Do_mS)_{ab}\Gamma^{ab}_{AB}+\coeff{i\sqrt2}{14} \,f\,e_{ma}\Gamma^a_{AB}-\coeff{\sqrt 2}{48}\,e_m{}^aF_{abcd}\Gamma^{bcd}_{AB}\,,\\[6 pt]
\cA_m{}_{ABCD} & \eql -\coeff34 \, (S^{-1} \Do_mS)_{ab}\Gamma^{a}_{[AB}\Gamma^b_{CD]}+\coeff{i\sqrt 2}{56}\, e_{ma}f\,\Gamma^{ab}_{[AB}\Gamma^b_{CD]} 
+\coeff{\sqrt 2}{32}\,e_m{}^aF_{abcd} \Gamma^{b}_{[AB}\Gamma^{cd}_{CD]}\,.\label{su8ABA}
\end{align}
The matrix $S_a{}^b$ was already defined in (\ref{S}), while the remaining flux 
terms originate from the four-form field strength of $d\!=\!11$ supergravity in the standard 
way. These two formulae are only valid in the `real vielbein gauge'  \eqref{genvield11}, 
and they will change when we switch to another SU(8) gauge. On the other hand, when
reverting from a general SU(8) covariant expression back to the real gauge, one must,
of course, ensure that the resulting expressions preserve the tensor structure
inherited from $d\!=\!11$ supergravity, which is manifest in \eqref{su8ABB} and 
\eqref{su8ABA}.\footnote{In addition, the flux components
  must satisfy the Bianchi identities. As shown in \cite{deWit:1986mz}, this 
  is guaranteed by the $\rm SU(8)$ covariant field equations.}

An important feature of the real gauge \eqref{genvield11}, not spelled out in 
\cite{deWit:1986mz}, is that (\ref{su8ABB}) is in fact {\em not} the most  general 
solution; rather, the {\it GVP\/} will still be satisfied if we replace
\begin{equation}\label{X}
f e_{ma} \;\rightarrow\; e_m{}^a X_{a|b} \, , \qquad
e_m{}^aF_{abcd} \;\rightarrow\; e_m{}^aX_{a|bcd} \,, 
\end{equation}
where $X_{a|b}$ is an arbitrary matrix, and $X_{a|bcd}$ is anti-symmetric only
in the indices $[bcd]$. In other words, the {\em GVP} admits solutions which
in general  are not compatible with the properties of the fluxes dictated by $d\!=\!11$ 
supergravity. A crucial requirement for the consistency of the truncation is therefore 
to ensure that the embedding formula for the fluxes respects these properties. We will call it the 
{\it correct tensor structure}  condition.

Let us now turn to the $d\!=\!4$ side of the story. Here we have formally the same 
{\it GVP\/} equation
\begin{equation}\label{vieleqs}
\Do_me^n_{\,ij}+\cals B_m{}^k{}_{[i}\,e^n_{\,j]k}+\cals A_{m\,ijkl}\,e^{n\,kl}\eql 0\,,
\end{equation}
but where the $\rm SU(8)$ gauge field $\cals B_{m}{}^i{}_j$  and the self-dual tensor field 
$\cals A_{m\,ijkl}$   are now to be expressed in terms of the $d\!=\!4$ fields. To match the
$d\!=\!11$ supersymmetry variations with those of gauged $\cN\!=\!8$ supergravity,
however, we must invoke a second consistency requirement.
This is to require that the $\frak A_1$ and $\frak A_2$ tensors, defined as \cite{deWit:1986iy}
\begin{equation}\label{deffAone}
\frak A_1^{ij}\eql -\coeff{\sqrt 2}{4}\,(e^{m\,ik}\,\cals B_m{}^j{}_k+\cals A_m{}^{ijkl}e^m_{kl})\,,
\end{equation}
\begin{equation}\label{deffAtwo}
 \frak A{}_{2l}{}^{ijk}\eql -\coeff{\sqrt 2} 4\,(3\, e^{m\,[ij}\cals B_m{}^{k]}{}_l-3\,e^m_{\,pq}\,\cals A_m{}^{pq[ij}\delta^{k]}{}_l-4 \,\cals A_m{}^{ijkp}e^m_{\,pl})\,,
\end{equation}
are equal to the corresponding  $A_1$ and $A_2$ tensors\footnote{Explicit formulae for $A_1$ and $A_2$ are given in \eqref{Ttensor} and \eqref{Atensors} 
below.}
 of $\cN\!=\!8$, $d\!=\!4$ supergravity,
which parametrize the $g$-dependent deformations (Yukawa couplings and scalar
potential) from the ungauged theory:
\begin{align}\label{frakAeqs1}
\frak A_1^{ij} & \eql g\, A_1^{ij}\,,\\ \label{frakAeqs2} 
\frak A_{2\,l}{}^{ijk} & \eql g\, A_{2\,l}{}^{ijk}\,.
\end{align}
In the remainder we will refer to these equations as the `$\frak A$-equations.'

The role of these two conditions in the consistent truncation is to ensure that the 
dependence  on the internal space drops out in the reduction of the supersymmetry 
variations of $d\!=\!11$ supergravity to four dimensions, and that in the process one 
recovers the complete supersymmetry transformations of gauged $\cN\!=\!8$ supergravity. 
In particular, one should note that while the $A_1$ and $A_2$ tensors on the right hand 
side in \eqref{frakAeqs1} and \eqref{frakAeqs2} are functions only of the scalar 56-bein 
of the $d\!=\!4$ theory, hence depend only on $x$, the  $\frak A_1$ and $\frak A_2$ tensors 
are hybrid objects that {\it a priori} depend both on the scalar 56-bein and the internal coordinates. 

A special solution of the {\em GVP\/} \eqref{vieleqs}  and the $\frak A$-equations \eqref{frakAeqs1} and \eqref{frakAeqs2} was constructed explicitly in  \cite{deWit:1986iy} in 
terms of the scalar 56-bein and the Killing vectors on $S^7$ as follows. Consider
\begin{equation}\label{theBzero}
\begin{split}
\cals B_m{}^i{}_j (\alpha,\beta)
& \eql -\coeff 2 3 \,\alpha\, m_7\,K_m{}^{IJ}(u^{ik}{}_{IK}u_{jk}{}^{JK}-v^{ikIK}v_{jkJK})\\
& \hspace{1in}
-\coeff 2 3\,\beta \Do_m K_{n}{}^{[IJ}K^{n\,KL]}(v^{ikIJ}u_{jk}{}^{KL}-u^{ik}{}_{IJ}v_{jkKL})\,,
\end{split}
\end{equation}
\begin{equation}\label{theAzero}
\begin{split}
\cals A_{m\,ijkl}(\alpha,\beta)
& \eql \alpha\,m_7\,\,K_m{}^{IJ}\,(v_{ijJK}u_{kl}{}^{IK}-u_{ij}{}^{JK}v_{klIK})\\
& \hspace{1in}-
\beta \Do_m K_{n}{}^{[IJ}K^{n\,KL]}(u_{ij}{}^{IJ}u_{kl}{}^{KL}-v_{ijIJ}v_{klKL})\,,
\end{split}
\end{equation}
where $\alpha$ and $\beta$ are arbitrary real parameters (recall that the indices 
on the Killing vectors are raised with the $S^7$ background metric, that is,
$K^{m\,IJ} \equiv \go^{mn} K_n^{IJ}$).
Substitution of these expressions into \eqref{vieleqs} shows that the 
{\it GVP\/} is satisfied provided $\alpha+4\beta=1$, 
leaving a one-parameter family of solutions. 

The remaining freedom is then fixed by imposing  the $\frak A$-equations. This is
by no means obvious, but happily, the detailed analysis of \cite{deWit:1986iy}  shows 
that the $\frak A$-equations do have a solution of the form above and 
indeed fix the free coefficients uniquely,
\begin{equation}\label{specalpha}
\alpha\eql\coeff 4 7\,,\qquad \beta \eql \coeff 3 {28}\,,
\end{equation}
when the gauge coupling constant of the $d\!=\!4$ theory is set to the inverse radius of $S^7$,
\begin{equation}\label{gtom7}
g\eql\sqrt 2 \,m_7\,.
\end{equation}
With these values, one re-obtains the correct four-dimensional
expressions, such that all dependence on the internal coordinates drops out
on the left hand side of \eqref{frakAeqs1} and \eqref{frakAeqs2}, as required 
for consistency. For this reason, we will refer to the solution \eqref{theBzero} and
\eqref{theAzero} {\em with the special values} \eqref{specalpha} as the {\em standard 
inhomogeneous solution} of the {\it GVP}, and simply denote it by
$(\cBo_m{}^i{}_j , \cAo_{m\, ijkl})$.

Nevertheless, direct translation of \eqref{theBzero} and \eqref{theAzero}
into the $d\!=\!11$ expressions (keeping track of the $\rm SU(8)$ alignment
rotation, see below) leads to apparent discrepancies with $d\!=\!11$ 
supergravity, in the sense that the resulting expressions in general will
{\em not} respect the tensor structure required by \eqref{su8ABB} and \eqref{su8ABA}.
The main new result of the present paper is to show how this defect 
can be remedied: namely, the {\it GVP} and the $\frak A$-equations still 
leave the freedom to modify the standard inhomogeneous solution by a homogeneous
term `in the kernel of the {\it GVP\/} and the $\frak A$-equations,'
and {\em this extra homogeneous contribution is precisely what is needed for
the fluxes in \eqref{su8ABB} and \eqref{su8ABA} to acquire the requisite 
tensor structure compatible with $d\!=\!11$ supergravity.}  We show that such a correction exists and is unique for {\em any} point on 
the scalar manifold  $\rm E_{7(7)}/SU(8)$.

Put another way,  given any solution to the {\em GVP}  \eqref{vieleqs} and the $\frak A$-equations  \eqref{frakAeqs1}  and \eqref{frakAeqs2}, viewed as  a system of linear equations for $\cals B_m{}^i{}_j$ and $\cals A_{m\,ijkl}$ on the $d\!=\!4$ side, one can, at least in principle,  determine the corresponding solution of the {\em GVP\/} \eqref{vieleqs11} on the $d\!=\!11$ side by performing the $\rm SU(8)$ gauge transformation:
\begin{equation}\label{Utransform}
\begin{split}
\cals B_m{}^A{}_B   & \eql U^A{}_i \big[ \cals B_m{}^i{}_j+2\Do_m \big]U^j{}_B\,,\\[6 pt]
\cals A_{m\,ABCD}  & \eql \cals A_{m\,ijkl}U^i{}_AU^j{}_BU^k{}_CU^l{}_D\,,
\end{split}
\end{equation}
where  
\begin{equation}
U^A{}_i(x,y) \equiv \Phi^A{}_B (x,y)\eta^B{}_i(y)\,,
\end{equation}
involves both the $\rm SU(8)$ alignment matrix $\Phi(x,y)$ and the conversion matrix $\eta(y)$
({\it alias} Killing spinor) between the two kinds of $\rm SU(8)$ indices. Note that
$\cals B_m$ transforms as a {\em bona fide} SU(8) gauge connection with the usual 
inhomogeneous contribution. There is {\it a priori} no reason why the standard 
inhomogenous solution \eqref{theAzero} and \eqref{theBzero} would yield the particular solution \eqref{su8ABB} and \eqref{su8ABA} with the correct tensor structure in $d\!=\!11$, and,  in fact, as we verify explicitly in Section~\ref{secsix}, in general it  does not.  
This means that the proof of the consistent truncation 
based on that particular solution is incomplete;  one must still show that for each point 
on the $\rm E_{7(7)}/SU(8)$ coset there exists a  solution to the linear system \eqref{vieleqs}, \eqref{frakAeqs1}  and  \eqref{frakAeqs2}  on the $d\!=\!4$ side that 
does have the correct tensor structure in $d\!=\!11$. 

Our strategy will be to look for 
a correction  $(\delta \cals B_m{}^i{}_j\,,\,\delta\cals A_{m\,ijkl})$ to the standard 
inhomogeneous solution and solving the homogeneous part of the {\it GVP}, such that
\begin{equation}\label{delBA}
\cals B_m{}^i{}_j\eql \cBo_m{}^i{}_j+\delta \cals B_m{}^i{}_j\quad,\qquad 
\cals A_{m\,ijkl}\eql \cAo_{m\,ijkl}+\delta \cals A_{m\,ijkl} \,,
\end{equation}
satisfy all consistency conditions. In particular, the `corrections' $\delta\cals B_m$ 
and $\delta \cals A_m$ must drop out of the $\frak A$-equations (hence belong
to their `kernel'), as otherwise the agreement with the $d\!=\!4$ theory
would be spoiled! Identifying these `corrections' is actually a simpler problem than finding 
a full solution because the tensors    $(\delta \cals B_m{}^i{}_j\,,\,\delta\cals A_{m\,ijkl})$ satisfy 
the homogenous system of equations corresponding to \eqref{vieleqs}, \eqref{frakAeqs1}  
and  \eqref{frakAeqs2} and transform covariantly under the $U$-rotation. We will  
not present a closed form solution (which is available in principle, but very cumbersome),
but we do prove that it always exists and is unique. Quite remarkably,
it will turn out that the closed form solution is not even needed to extract 
the non-linear ans\"atze for the fluxes!

Given a solution to all consistency conditions, the fluxes can be either read
off from the expansions \eqref{su8ABB} and/or \eqref{su8ABA}, or alternatively 
from the $\rm SU(8)$-invariant projection that is summarized in the flux formula (7.5) 
in \cite{deWit:1986iy}~\footnote{We correct some typos in the original formula. For clarity
  of notation we put a bar on the complex conjugate vielbein $e^{mij}\equiv \bar e^{mij}$
  in the formulae below, whenever the $\rm SU(8)$ indices are not written out.}
\begin{equation}\label{fluxeqs}
\coeff 4 7 \,if\,g_{n[p}\delta_{q]}{}^m+\coeff 1 2 \, F^m{}_{npq}\eql { -} i\coeff {\sqrt 2}{480}\,\Delta^4\,\epsilon_{pqrstuv}\,e^m_{ij}(e^r\bar e^se^t\bar e^ue^v)_{kl}\cals A_n{}^{ijkl}\,.
\end{equation}
Let us emphasize once again that in general this equation is inconsistent if 
$\cals A_{m}{}^{ijkl}$ is replaced with the standard one ($= \!\cAo_{m}{}^{ijkl}\,$) but, as 
we will show, there always exists a unique $\delta \cals A_{m\,ijkl}$ such that \eqref{fluxeqs}
does hold with \eqref{delBA}. Solving \eqref{fluxeqs} for the flux components, we find
\begin{equation}\label{thefproj}
f\eql -\coeff {\sqrt 2}{48\cdot 5!}\,\Delta^4\,g^{mu}\,\epsilon_{mnpqrst}\,e^n_{ij}\,(e^{[p}\bar e^q e^r\bar e^s e^{t]})_{kl}\,\cals A_u{}{}^{ijkl}\,,
\end{equation}
and
\begin{equation}\label{theFproj}
F_{mnpq}\eql -\coeff{i}{144}\,\Delta^4\,g_{rw}\,e^r_{ij}(e^{[s}\be^te^u\be^ve^{w]})_{kl}\,\epsilon_{stuw[mnp}\,\cals A_{q]}{}^{ijkl}\,.
\end{equation}
Note that in order to exploit these equations we must first solve for the full metric $g_{mn}$
from \eqref{metrivfor}. As we will explain in detail below, $f$ and $F_{mnpq}$ as given in 
\eqref{thefproj} and \eqref{theFproj} are {\em invariants of the linear system}. 
In other words, it does not matter which solution to the linear system 
\eqref{vieleqs}, \eqref{frakAeqs1}  and  \eqref{frakAeqs2} one uses to evaluate them by 
projecting the right hand side of \eqref{fluxeqs} onto the components \eqref{thefproj}
and \eqref{theFproj}: all solutions, in particular the standard inhomogeneous solution 
\eqref{theBzero}-\eqref{specalpha} give the same answer! In this sense \cite{deWit:1986iy} 
contains already the complete result for the fluxes. Besides the analytic examples of
Section~\ref{secsix} we will present some non-trivial numerical checks of \eqref{thefproj}
in Section~\ref{secseven}.

The rest of this paper is more technical, as we present the details of our calculations. 
In Section \ref{sectwo} we determine the general solutions for {\it GVP\/}  
\eqref{vieleqs} and \eqref{vieleqs11}. The $\frak A$-equations are included in  
Section \ref{secthree}. We show that the full linear system is invariant under the 
natural action of $\rm E_{7(7)}$ and use this symmetry to determine the general solution for all consistency conditions on the $d\!=\!4$ and the $d\!=\!11$ sides. This allows us to   complete the proof of the consistent truncation. Explicit flux formulae are rederived in Section \ref{secfive}. In Section~\ref{secsix} 
we illustrate various point of the construction on two examples, the $\rm SO(7)^-$ and $\rm SO(7)^+$ families. Some results of numerical explorations are summarized in Section~\ref{secseven} and we conlude   in Section~\ref{seceight}.

\section{The Generalized Vielbein Postulate (GVP)}
\label{sectwo}
\setcounter{equation}{0}

We now return to the vielbein equation \eqref{vieleqs} in order to explain the construction 
in more detail and to work out the most general solution of the {\em GVP}.
For  a given vielbein,   \eqref{vieleqs}  can be viewed as an inhomogenous linear 
equation for the 
components of the $\rm SU(8)$ gauge field, $\cals B_m{}^i{}_j$, and the tensor 
field, $\cals A_{m\,ijkl}$. Recall that the generalized vielbein, $e^m_{\,ij}$, and its complex conjugate, $e^{m\,ij}$, can be assigned to transform in the $\bf 56$-dimensional 
representation of $\rm E_{7(7)}$. For a given point on the scalar coset represented by the 
group element $\cals V(x)$, cf. \eqref{V}, we can rewrite  \eqref{genvieldef} as
\begin{equation}\label{vieltrans}
\left(\begin{matrix}
e^m_{\,ij}\\ e^{m\,ij} 
\end{matrix}\right)\eql 
\left(\begin{matrix}
u_{ij}{}^{IJ} & v_{ijIJ}\\ v^{ijIJ} & u^{ij}{}_{IJ}
\end{matrix}\right)
\left(\begin{matrix}
K^{m}{}_{IJ}\\ K^{m\,IJ}
\end{matrix}\right)\,,
\end{equation}
where $K^m_{\,IJ}\equiv K^{m\,IJ}$, as the Killing vectors are real. Similarly, 
$\cals B_m{}^i{}_j$ and $\cals A_{m\,ijkl}$, together  with the complex conjugates, 
can be assigned to the adjoint representation $\bf{133}$  of $\rm E_{7(7)}$, 
\begin{equation}\label{Eseventr}
\left(\begin{matrix}
\delta_{[i}{}^{[k}\cals B_{m\,j]}{}^{l]} & \cals A_{m\,ijkl}\\ \cals A_{m}{}^{ijkl} & \delta^{[i}{}_{[k}\cals B_m{}^{j]}{}_{l]}\end{matrix}\right)
\eql \cals V \,
\left(\begin{matrix}
\delta_{[I}{}^{[K}   B_{m\,J]}{}^{L]} &   A_{m\,IJKL}\\   A_{m}{}^{IJKL} & \delta^{[I}{}_{[K}   B_m{}^{J]}{}_{L]}
\end{matrix}\right)\,\cals V^{-1}\,.
\end{equation}
By performing a purely $x$-dependent  $\rm E_{7(7)}$ rotation\footnote{Which from
  the $d\!=\!7$ perspective looks like a {\em rigid} transformation.} by $\cals V^{-1}(x)$, one transforms the  {\it GVP} \eqref{vieleqs} into an equation that has no explicit 
dependence on the scalar 56-bein any more \cite{deWit:1986iy},
\begin{equation}\label{Kvieleqs}
\Do_m\,K^n_{\,IJ}+     B_m{}^K{}_{[I}K^n_{\,J]K}+  A_{m\,IJKL}K^{n\,KL}\eql 0\,.
\end{equation}
This equation can be further simplified if we consider the Killing spinors $\eta^I{}_A$ in \eqref{Kvectdef} as a local $\rm SO(8)\subset SU(8)$ transformation on $S^7$. Taking into account the inhomogeneous term in the transformation of $  B_m{}^I{}_J$, cf.\ \eqref{Utransform}, we can also remove the explicit dependence on the $S^7$ coordinates in \eqref{Kvieleqs}, 
\begin{equation}\label{homvieleqs}
  B_m{}^C{}_{[A}\Gamma^n_{\,B]C}+ A_{m\,ABCD}\Gamma^{n}_{\,CD}\eql 0\,,
\end{equation}
ending up with  a homogenous equation for $B_m{}^A{}_B$ and $A_{m\,ABCD}$. This amounts to analyzing \eqref{Kvieleqs}  at the North Pole of $S^7$, from
where it can be parallel transported back to any other point by application of the 
matrix $\eta$. The anti-hermitean $ B_{m}{}^A{}_B$ and the complex-selfdual 
$A_{m\,ABCD}$ can now be expanded into a 
basis of $\Gamma$-matrices as follows: 
\begin{equation}\label{Btens}
\begin{split}
 B_m{}^A{}_B\eql \alpha_{m\,ab}\,\Gamma^{ab}_{AB}-i\,\alpha_{m\,a}\,\Gamma^a_{AB}+\alpha_{m\,abc}\,\Gamma^{abc}_{AB}\,,
\end{split}
\end{equation}
\begin{equation}\label{Atens}
A_{m\,ABCD}\eql \beta_{m\,ab}\,\Gamma^a_{[AB}\Gamma^b_{CD]}+i\,\beta_{m\,a}\,\Gamma^{ab}_{[AB}\Gamma^b_{CD]}+\beta_{m\,abc}\,\Gamma^{[a}_{[AB}\Gamma^{bc]}_{CD]}\,,
\end{equation}
where the expansion coefficients  are (anti-)symmetric according to  the contraction with the $\Gamma$-matrices, but otherwise real and arbitrary. Substituting those 
expansions into \eqref{homvieleqs},  we obtain an equation that is antisymmetric in the spinor indices. The contraction  with $\Gamma^a_{AB}$ and $\Gamma^{ab}_{AB}$ projects out two independent equations for the expansion coefficients:
\begin{equation}\label{veqshone}
\alpha_{m\,ab}+\coeff 2 3\,\beta_{m\,ab}+\coeff 1 3\,\delta_{ab}\,\beta_{m\,cc}\eql 0\,,
\end{equation}
\begin{equation}\label{veqshtwo}
\alpha_{m\,abc}+\coeff 2 3\,\beta_{m\,abc}-\coeff i 3\, \delta_{a[b}\,(\alpha_{m\,c]}+4\beta_{m\,c]})\eql 0\,,
\end{equation}
where the antisymmetrization is only over the flat indices.  From the symmetric and antisymmetric parts of \eqref{veqshone} and the real and imaginary parts of \eqref{veqshtwo}, we obtain the general solution:
\begin{equation}\label{alphabetaone}
\alpha_{m\,ab}\eql 0\,,\qquad \beta_{m\,ab}\eql 0\,,
\end{equation}
\begin{equation}\label{alphabetatwo}
\alpha_{m\,a}+4\beta_{m\,a}\eql 0\,,\qquad \alpha_{m\,abc}+\coeff 2 3\, \beta_{m\,abc}\eql 0\,.
\end{equation}
Rotating back   with Killing spinors gives the general solution to \eqref{Kvieleqs}, which may be partially recast in terms of the Killing vectors:
\begin{equation}\label{Bkilling}
\begin{split}
B_m{}^I{}_J & \eql -i\,(m_7\,\eo_{ma}+\alpha_{m\,a})\,\bar\eta^I\Gamma^a\eta^J+\alpha_{m\,abc}\,\bar\eta^I\Gamma^{abc}\eta^J\\
&\eql  -\,(m_7\,\delta_m{}^n+\alpha_{m}{}^{n})\, K_n{}^{IJ}+\alpha_{m\,abc}\,\bar\eta^I\Gamma^{abc}\eta^J\,,
\end{split}
\end{equation}
\begin{equation}\label{Akilling}
\begin{split}
A_{m\,IJKL} & \eql i\,\beta_{m\,a}\,\bar\eta^{[I}\Gamma^{ab}\eta^{J}\bar\eta^K\Gamma^b\eta^{L]}+
\beta_{m\,abc}\,\bar\eta^{[I}\Gamma^{ab}\eta^{J}\bar\eta^K\Gamma^{c}\eta^{L]}\\
 & \eql -\,\beta_{m}{}^n\,\Do_nK_p^{[IJ}K^{p\,IJ]}+
\beta_{m\,abc}\,\bar\eta^{[I}\Gamma^{ab}\eta^{J}\bar\eta^K\Gamma^{c}\eta^{L]}\,,
\end{split}
\end{equation}
where 
\begin{equation}\label{}
\alpha_m{}^n\eql\alpha_{m\,a}\eo^{an}\,,\qquad \beta_m{}^n\eql\beta_{m\,a}\eo^{an}\,.
\end{equation}
Finally, by applying the $\rm E_{7(7)}$ transformation \eqref{Eseventr} to \eqref{Bkilling} and \eqref{Akilling}, we obtain the general solution to the {\it GVP\/} \eqref{vieleqs} on the $d=4$ side.

We see that the one parameter family in the standard solution \eqref{theBzero} and \eqref{theAzero} of the vielbein equation corresponds to the special choice
\begin{equation}\label{getalpgha}
\alpha_m{}^n\eql (\alpha-1)\,m_7\,\delta_m{}^n\,,\qquad \beta_m{}^n\eql \beta\,\delta_m{}^n\,,\qquad \alpha+4\beta\eql 1\,.
\end{equation}
This family is distinguished in that $\cals B_m{}^i{}_j$ and $\cals A_{m\,ijkl}$ are constructed entirely from the 56-bein and the Killing vectors, and are `covariant'  with respect to the round $S^7$ in the sense that $\alpha_{m\,a}\sim\beta_{m\,a}\sim \eo_{ma}$. However, we will see in the following 
that the consistent truncation calls for more general solutions than the standard one.

To enumerate all solutions to the {\em GVP\/} in $d\!=\!4$, we introduce the independent parameters with  flat indices,
\begin{equation}\label{}
\alpha_{a|b}\eql \eo_a{}^m\,\alpha_{m\,b}\,,\qquad \alpha_{a|bcd}\eql \eo_a{}^m\,\alpha_{m\,bcd}\,,
\end{equation}
which, as suggested by the notation,  have no a priori symmetry between the first index 
and the remaining ones. Hence, under the $\rm SO(7)$ tangent rotations acting on the 
{\it background\/} siebenbein~$\eo^a$, they decompose 
into the following irreducible components:
\begin{equation}\label{Xfour}
\begin{split}
\alpha_{a|b} & \eql \alpha^{\,\Yboxdim4pt\yng(1,1)}_{ab}+ \alpha^{\,\Yboxdim4pt\yng(2)}_{ab} + 
(\alpha-1)m_7\, \delta_{ab} \,\,,
\\[6 pt]
\alpha_{a|bcd} & \eql \alpha_{abcd}^{\,\Yboxdim4pt\yng(1,1,1,1)}+\alpha_{abcd}^{\,\Yboxdim4pt\yng(2,1,1)}+
\delta^{\phantom{{\,\Yboxdim4pt\yng(1,1)}}}_{a[b}\tilde\alpha_{cd]}^{\,\Yboxdim4pt\yng(1,1)}\,,
\end{split}
\end{equation}
all of which will in general be present in any particular solution (note that $\alpha_{ab}$
and $\tilde\alpha_{ab}$ are different).

Let us now turn to the  {\it GVP\/} \eqref{vieleqs11} on the $d\!=\!11$ side. Using covariance of the {\it GVP\/}, the $\rm SU(8)$ gauge field $\cals B_m{}^A{}_B$ and the self-dual tensor $\cals A_{m\,ABCD}$ of the general solution in $d\!=\!11$ can be obtained by applying the $\rm SU(8)$ gauge transformation (the $U$-rotation) \eqref{Utransform} to the general solution for  $\cals B_m{}^i{}_j$ and $\cals A_{m\,ijkl}$ in \eqref{Bkilling} and \eqref{Akilling}, respectively. However, since the $U$-rotation is not known explicitly, we will proceed differently and solve \eqref{vieleqs11} directly on the $d\!=\!11$ side. 

First we note that the decomposition of $\cals B_m{}^A{}_B$ and $\cals A_{m\,ABCD}$ into irreducible $\rm SU(8)$ components is given by   precisely  the same expansions as in \eqref{Btens} and \eqref{Atens}, respectively,  although  values of the expansion parameters for a given solution will in general be different on the $d\!=\!4$ and the $d\!=\!11$ side. Next we evaluate the derivative of the generalized vielbein  in \eqref{vieleqs11}, which, using \eqref{genvield11} and \eqref{S}, can be expressed in terms of the matrix $S^{-1}\Do_m S$ \cite{deWit:1986mz}.  It is then straightforward to check that the first terms 
in \eqref{su8ABB} and \eqref{su8ABA} given by the  antisymmetric and symmetric parts 
of $(S^{-1}\Do_mS)_{ab}$, respectively, already solve \eqref{vieleqs11} by themselves and that the resulting homogeneous equation that determines 
parameters of the general solution is exactly the same as  \eqref{homvieleqs} or, equivalently, \eqref{alphabetaone} and \eqref{alphabetatwo}. Hence, we may readily write down the most general solution  to \eqref{vieleqs11} which, in accordance with \eqref{X}, is given by 
\begin{equation}\label{Bd11}
\cals B_m{}^A{}_B   \eql\coeff 1 2 \,(S^{-1} \Do_mS)_{ab}\Gamma^a_{AB}+4 i\,X_{m\,a}\,\Gamma^a_{AB}-\coeff2 3\,X_{m\,bcd}\,\Gamma^{bcd}_{AB}\,,
\end{equation}
\begin{equation}\label{Ad11}
\begin{split}
\cA_m{}_{ABCD} & \eql -\coeff34 \, (S^{-1} \Do_mS)_{ab}\Gamma^{a}_{[AB}\Gamma^b_{CD]}+i\,X_{m\,a}\,\Gamma^b_{[AB}\Gamma^{ab}_{CD]}+
 X_{m\,bcd}\,\Gamma^{b}_{[AB}\Gamma^{cd}_{CD]}\,,
\end{split}
\end{equation}
where $X_{m\,a}$ and $X_{m\,abc}=X_{m\,[abc]}$ are real and otherwise arbitrary.\footnote{We also rescaled them with respect to  \eqref{X}.}

After conversion to flat indices, the tensors
\begin{equation}\label{Xtwo}
X_{a|b}\eql e_a{}^m\,X_{m\,b}\,,\qquad X_{a|bcd}\eql e_a{}^m\,X_{m\,bcd}\,,
\end{equation}
may be decomposed into  irreducible components under the $\rm SO(7)$ rotations acting on the metric  vielbein $e^a$, 
\begin{equation}\label{XXfour}
\begin{split}
X_{a|b}& \eql X^{\,\Yboxdim4pt\yng(2)}_{ab} + X^{\,\Yboxdim4pt\yng(1,1)}_{ab}+\delta_{ab}\,X\,,\\[10 pt]
X_{a|bcd}& \eql X_{abcd}^{\,\Yboxdim4pt\yng(1,1,1,1)}+X_{abcd}^{\,\Yboxdim4pt\yng(2,1,1)}+
\delta_{a[b}^{\phantom{\Yboxdim4pt\yng(1,1)}}\tilde X_{cd]}^{\,\Yboxdim4pt\yng(1,1)}\,.
\end{split}\end{equation}
with the same representation content as in \eqref{Xfour}. 

Comparing with the solution  \eqref{su8ABB} and \eqref{su8ABA}, we see that the only $\rm SO(7)$ representations in \eqref{Xtwo} and \eqref{Xfour} that are consistent with the supersymmetry in $d\!=\!11$ are the singlet  and the totally antisymmetric one. Those are determined by the components of the flux, 
\begin{equation}\label{specfsol}
X\eql \coeff{\sqrt 2}{56}\, f\,, \qquad X_{abcd}^{\,\Yboxdim4pt\yng(1,1,1,1)}\eql \coeff{\sqrt 2}{32}F_{abcd}^{\phantom{\Yboxdim4pt\yng(1,1,1,1)}}\,.
\end{equation}
The correct tensor structure condition simply means that the $X$-parameters in all other representations must vanish.

\begin{table}[t]
\begin{center}
\scalebox{0.9}{
\begin{tabular}{@{\extracolsep{25 pt}}c c c c l}
\toprule
\noalign{\smallskip}
GV & $\rm E_{7(7)}$ connection & Parameters & GVP & Rotation\\
\noalign{\smallskip}
\midrule
\noalign{\smallskip}
$i\,\Gamma^m_{AB}$ & $B_m{}^A{}_B\,,\, A_{m\,ABCD}$  & $\alpha_{a|b}\,,\alpha_{a|bcd}$ &  {\small \eqref{homvieleqs}}  \\
\noalign{\vspace{-6 pt}}
& & & & $\eta\in{\rm SO(8)}$\\
\noalign{\vspace{-6 pt}}
$K^m_{\,IJ}$ & $ B_{m}{}^I{}_J\,,\, A_{m\,IJKL}$ & $\alpha_{a|b}\,,\alpha_{a|bcd}$ &  {\small \eqref{Kvieleqs}}  \\
\noalign{\vspace{-6 pt}}
& & & & $\cals V\in{\rm E_{7(7)}}$\\
\noalign{\vspace{-6 pt}}
$ e^m_{\,ij}$ & $\cals B_m{}^i{}_j\,,\,\cals A_{m\,ijkl} $ & $\alpha_{a|b}\,,\alpha_{a|bcd}$ &  {\small \eqref{vieleqs}} \\
\noalign{\vspace{-6 pt}}
& & & & $  U\in{\rm SU(8)}$ \\
\noalign{\vspace{-6 pt}}
$e^m_{AB}$ & $\cals B_m{}^A{}_B\,,\,\cals A_{m\,ABCD}$ & $X_{a|b}\,,X_{a|bcd}$ &  {\small \eqref{vieleqs11}} \\[5 pt]
\bottomrule
\end{tabular}
}
\caption{\label{thevielflow}
Generalized Vielbein Postulate in different frames.  
}
\end{center}
\end{table}

In Table~\ref{thevielflow} we have summarized the different forms of  the vielbein equation 
that we looked at in this section and listed the functions that parametrize the space of 
solutions. While it is clear that any solution on the $d\!=\!4$ side  given in terms of 
$\alpha_{a|b}(x,y)$ and $\alpha_{a|bcd}(x,y)$ maps under the $\rm SU(8)$ rotation, 
$U(x,y)$,  onto a unique solution  given in terms of $S(x,y)$, $X_{a|b}(x,y)$ and 
$X_{a|bcd}(x,y)$, it is by no means guaranteed that the latter will be consistent with the 
tensor structure \eqref{specfsol} required by $d\!=\!11$ supergravity.

\section{The $\frak A$-equations}
\label{secthree}
\setcounter{equation}{0}

To resolve the possible remaining discrepancies we need to take a closer look at 
$\frak A$-equations \eqref{frakAeqs1} and \eqref{frakAeqs2} which, together with 
the {\it GVP} \eqref{vieleqs}, guarantee the consistent reduction of the supersymmetry transformations from  $d\!=\!11$  to $d\!=\!4$ \cite{deWit:1986iy}. The new key insight
of the present work is that these equations admit more general solutions than the one
given in \cite{deWit:1986iy}, and encapsulated in the standard inhomogeneous solution
\eqref{theBzero}-\eqref{specalpha}.

We first show that the system of vielbein equations and $\frak A$-equations is invariant 
under the action of $\rm E_{7(7)}$. To this end we recall
that the $A_1^{ij}$ and $A_{2\,l}{}^{ijk}$  tensors  and their complex conjugates
correspond to two $\rm SU(8)$ irreducible components of the so-called $T$-tensor 
of $\cN\!=\!8$, $d\!=\!4$ gauged supergravity. The latter is defined in terms of the 
scalar 56-bein \cite{de Wit:1982ig}
\begin{equation}\label{Ttensor}
T_i{}^{jkl}\eql (u^{kl}{}_{IJ}+v^{klIJ})(u_{im}{}^{JK}u^{jm}{}_{KI}-v_{imJK}v^{jmKI})\,.
\end{equation}
Then
\begin{equation}\label{Atensors}
A_1^{ij}\eql \coeff 4 {21}\, T_k{}^{ikj}\,,\qquad A_{2\,l}{}^{ijk}\eql -\coeff 4 3\,T_l{}^{[ijk]}\,.
\end{equation}

It follows from the variations of the $T$-tensor in \cite{de Wit:1982ig} that under infinitesimal transformations of the scalar vielbein,
\begin{equation}\label{}
\delta\cals V
\eql 
\left(\begin{matrix}
\delta_{[i}{}^{[k}\Lambda_{j]}{}^{l]} & \Sigma_{ijkl} \\ 
\Sigma^{ijkl} & \delta^{[i}{}_{[k}\Lambda^{j]}{}_{l]}
\end{matrix}\right) \,\cals V \,,
\end{equation}
where $\Lambda^i{}_j$ are antihermitian and $\Sigma_{ijkl}$ are self-dual, 
the $A_1^{ij}$ and $A_{2\,i}{}^{ijkl}$ tensors together transform in the $\bf 912$ irreducible 
representation of $\rm E_{7(7)}$ \cite{deWit:1983gs}.~\footnote{The modern
  formulation of gauged supergravities relies on the {\em embedding tensor formalism}
  \cite{NicSamt,dWST}. The above transformation property of the $T$-tensor then simply
  expresses the so-called {\em representation constraint} that the embedding tensor 
  must satisfy.}
We will now show that the $E_{7(7)}$ transformations 
of the generalized vielbein and the tensor fields in  \eqref{vieltrans} and \eqref{Eseventr} 
induce exactly the same $\rm E_{7(7)}$ transformations of the  composite $\frak A_1^{ij}$ and $\frak A_{2\,i}{}^{jkl}$ tensors defined in \eqref{deffAone} and \eqref{deffAtwo}. 

Since the   $\rm SU(8)$ covariance is manifest,   all we must show is that under 
infinitesimal transformation by the coset generators \cite{deWit:1983gs}
\begin{equation}\label{tranfAone}
\delta\frak A_1^{ij}\eql -\coeff 1 6\,(\frak A_2{}^i{}_{pqr}\Sigma^{jpqr}+\frak A_2{}^j{}_{pqr}\Sigma^{ipqr}
)\,,
\end{equation}
\begin{equation}\label{tranfAtwo}
\delta \frak A_{2\,i}{}^{jkl}\eql -2\,\frak A_{1 ip}\,\Sigma^{pjkl}-3\,\Sigma^{pq[jk}\,\frak A_{2}{}^{l]}{}_{ipq}- \Sigma^{pqr[j}\,\delta^k{}_i\,\frak A_2{}^{l]}{}_{pqr}\,.
\end{equation}
Evaluating  $\delta\frak A_1^{ij}$ from the definition \eqref{deffAone} and  setting it equal to  \eqref{tranfAone} gives
\begin{equation}\label{thevarAone}
\begin{split}
\Sigma^{ijkl}\,\cals B_m{}^p{}_{[k}\,e^m_{\,l]p}+\Sigma_{klpq}\,\cals A_m{}^{ijpq}\,e^{m\,kl}
+\coeff 2 3\,\Sigma_{klpq}\,\cals A_m{}^{jklp}\,e^{m\,iq}+\coeff 2 3\,\Sigma^{iklp}\,\cals A_{m\,klpq}\,e^{m\,jq}~\mathop{=}^?~ 0\,.
\end{split}
\end{equation}
Then using selfduality of $\Sigma_{ijkl}$ and $\cals A_{m\,ijkl}$, we can rewrite the 
second and the third terms as
\begin{equation}
\begin{split}
\Sigma_{klpq}\,\cals A_m{}^{ijpq}\,e^{m\,kl} & \eql \Sigma^{ijkl}\,\cals A_{m\,klpq}\,e^{m\,pq}+\coeff 1 6\,\Sigma^{klpq}\,\cals A_{m\,klpq}\,e^{m\,ij}\\[6 pt]
& \hspace{0.5in}-\coeff 2 3\,(\Sigma^{iklp}\,\cals A_{m\,klpq}\,e^{m\,jq}-\Sigma^{jklp}\,\cals A_{m\,klpq}\,e^{m\,iq})\,,\end{split}
\end{equation}
and
\begin{equation}\label{}
\begin{split}
\coeff 2 3\,\Sigma_{klpq}\,\cals A_m{}^{jklp}\,e^{m\,iq}\eql -\coeff 1 6\,\Sigma^{klpq}\,\cals A_{m\,klpq}\,e^{m\,ij}-\coeff 2 3\,\Sigma^{jklp}\,\cals A_{m\,klpq}\,e^{m\,iq}\,,
\end{split}
\end{equation}
respectively. This reduces \eqref{thevarAone} to the vielbein equation,
\begin{equation}\label{}
\Sigma^{ijkl}\,(\cals B_m{}^p{}_{[k}\,e^m_{\,l]p}+\cals A_{m\,klpq}\,e^{m\,pq})\eql 0\,,
\end{equation}
where we have used that
$\Do_me^m_{\,ij}\eql 0\,$,
which follows from the definition of the generalized vielbein in terms of the Killing vectors on $S^7$.

One can also check  the $\delta\frak A_1^{ij}$  equation \eqref{tranfAone} starting from the equivalent definition \cite{deWit:1986iy}
\begin{equation}\label{atwoidtopr}
\frak A_1^{ij}\eql \coeff{\sqrt 2}{4}\,e^{m\,k(i}\cals B{}_m{}^{j)}{}_{k}\,,
\end{equation}
which makes the symmetry in $(ij)$ manifest. Here, a simple substitution of variations yields the condition
\begin{equation}\label{}
\Sigma^{ipqr}\,\cals A_{m\,pqrs}\,e^{m\,js}+(i\leftrightarrow j)
~\mathop{=}^?~ 0\,.
\end{equation}
{Denote  $U^j{}_{pqr}=\cals A_{m\,pqrs}\,e^{m\,js}$ and use the self-duality of the $\rm E_{7(7)}$ generator to rewrite the left hand side in \eqref{atwoidtopr} as
\begin{equation}\label{}
\begin{split}
\coeff 1 {24} \,\Sigma_{xywz}\,\epsilon^{xywzipqr}\,U^j{}_{pqr}+(i\leftrightarrow j)\eql 
\coeff 1 {24} \,\Sigma_{xywz}\,\epsilon^{ixywzpqr}\,\delta^s{}_p\,U^j{}_{sqr}+(i\leftrightarrow j)\,.
\end{split}
\end{equation}
The vanishing of this expression follows now from the Schouten identity applied to the indices $xywzpqrsj$ and the explicit symmetry in $(ij)$.
}

The $\delta\frak A_{2i}{}^{jkl}$ equation obtained by comparing the variation of 
\eqref{deffAtwo} with \eqref{tranfAtwo}, is satisifed after using the vielbein equation and the selfduality of $\Sigma_{ijkl}$ and $\cals A_{m\,ijkl}$. The intermediate expressions are  more involved and we omit them here. 

The $\rm E_{7(7)} $ invariance of the $\frak A$-equations \eqref{frakAeqs1} and \eqref{frakAeqs2} on the space of solutions of the generalized vielbein equation \eqref{vieleqs} allows us to solve those equations at the origin of the coset, where
\begin{equation}\label{Asorig}
A_1^{ij}\big|_{\cV = {\bf 1}}\eql\delta_{ij}\,,\qquad A_{2\,i}{}^{jkl} \big|_{\cV = {\bf 1}} \eql 0\,,
\end{equation}
and  $\cals B_m{}^i{}_j$ and $\cals A_{m\,ijkl}$ are given in  \eqref{Bkilling} and \eqref{Akilling}. Furthermore, since the Killing spinors form an $\rm SO(8)$ matrix, they can be rotated away from all equations, which then involve only the parameters $\alpha_{a|b}$ and $\alpha_{a|bcd}$, and the $\Gamma$-matrices.

Let us first look at the $\frak A_1$ equation \eqref{frakAeqs1}. Using \eqref{Bkilling}, \eqref{atwoidtopr} and \eqref{Asorig} it becomes
\begin{equation}\label{A1eqsor}
g\,\delta_{ij}\eql \coeff{\sqrt 2}4\,i\,\Gamma^a_{k(i}\Big[-i\,(m_7\,\delta_{ab}+\alpha_{a|b})\Gamma^b_{j)k}
+\alpha_{a|bcd}\,\Gamma^{bcd}_{j)k}\Big]\,.
\end{equation}
Collecting the independent terms and using \eqref{gtom7}, we get
\begin{equation}\label{A2solution}
(3\,m_7+\alpha_{a|a})\,\delta_{ij}-i\,\alpha_{a|bcd}\,\Gamma^{abcd}_{ij}\eql 0\,,
\end{equation}
which sets the completely antisymmetric component  $\alpha_{[a|bcd]}$ to zero and
fixes the trace of $\alpha_{a|b}$ via \eqref{getalpgha} such that $\alpha =\frac47$.
All the other components of $\alpha_{a|b}$ and $\alpha_{a|bcd}$ are left arbitrary.

It is more tedious to check that the  $\frak A_2$-equation \eqref{frakAeqs2}, which 
now simply reads 
\begin{equation}\label{A2eqss}
\frak A_{2\,i}{}^{jkl} \big|_{\cV = {\bf 1}}=0\,,
\end{equation}
is also solved if \eqref{A2solution} is satisfied. 
To check this explicitly, we note that \eqref{A2eqss} is antisymmetric in $[jkl]$, hence  we may instead show that the two equations obtained by contracting \eqref{A2eqss} with $\Gamma^a_{jk}$ and $\Gamma^{ab}_{jk}$ are satisfied. 
This is actually easier than working with the original equation, which is a fourth-rank 
tensor that must be expanded in the basis of independent products of $\Gamma$-matrices. Instead, after the contraction, one ends up with a second rank tensor, which is much simpler. 
Still given all  anti-symmetrizations in \eqref{deffAtwo} and in \eqref{Akilling}, one ends 
up with  a large number of terms which are best handled by a computer.
 
To summarize, we have shown that a general  solution to the generalized vielbein equation and the $\frak A$-equations is given by
\begin{equation}\label{thed4sol}
\alpha_{a|b}\eql -\coeff 3 7 \,m_7\,\delta_{ab}+\alpha^{\,\Yboxdim4pt\yng(2)}_{ab}+\alpha^{\,\Yboxdim4pt\yng(1,1)}_{ab}\,,\qquad 
\alpha_{a|bcd}\eql \alpha^{\,\Yboxdim4pt\yng(2,1,1)}_{abcd}+
\delta_{a[b}^{\phantom{\Yboxdim4pt\yng(1,1)}} \tilde\alpha^{\,\Yboxdim4pt\yng(1,1)}_{cd]}\,,
\end{equation}
where we used the same notation as in \eqref{Xfour}. All parameters in \eqref{thed4sol} are completely arbitrary. The standard solution \eqref{theBzero}-\eqref{specalpha} is obtained by setting all those parameters to zero. Then from \eqref{getalpgha} we get  $\alpha=\coeff 4 7 $ which agrees with \eqref{specalpha}.

The reader might have noticed that the $\rm SO(7)$ representations that arise in the solution \eqref{thed4sol} are precisely the same representations that must be set to zero in the parameters  \eqref{Xtwo} of the general solution \eqref{Bd11} and \eqref{Ad11} to the {\it GVP}  in $d\!=\!11$  to obtain  standard form with the fluxes in \eqref{specfsol}. However, one must be careful here because,  as we have noted at the end of Section~\ref{sectwo}, the relation between the parameters $\alpha_{a|b}$ and $\alpha_{a|bcd}$ on the $d\!=\!4$ side and the parameters $X_{a|b}$ and $X_{a|bcd}$ on the $d\!=\!11$ side is by no means straightforward as it involves, see Table~\ref{thevielflow}, both the $\rm E_{7(7)}$ and  the $\rm SU(8)$ rotations that can mix different $\rm SO(7)$ components. 

Let us now choose a particular solution in $d\!=\!4$, for example the standard solution, $(\cBo_{m}{}^i{}_j\,,\cAo_{ijkl})$. Any other solution is then obtained by adding to it a solution, $(\delta\cB_m{}^i{}_j,\delta\cA_{m\,ijkl})$, to the homogenous equations given by setting the first term in \eqref{vieleqs} and the left hand sides in \eqref{frakAeqs1} and \eqref{frakAeqs2} to zero. Since $\delta\cB_m{}^i{}_j$ and $\delta\cals A_{m\,ijkl}$ transform homogeneously under $\rm SU(8)$, we may now apply the $U$ rotation to those homogeneous equations to revert to the real vielbein
gauge \eqref{genvield11}. After converting to flat indices using the metric vielbein, $e_a{}^m$, 
and dropping the warp factor, the homogeneous part of the {\it GVP} reduces to
\begin{equation}\label{}
\delta\cals B_a{}^C{}_{[A}\Gamma^b_{B]C}+\delta\cals A_{a\,ABCD}\Gamma^{a}_{CD}\eql 0\,,
\end{equation}
while the conditions for $(\delta\cals B_a{}^A{}_B , \delta\cals A_{a\, ABCD})$ 
to be `in the kernel of the $\frak A$-equations' become
\begin{equation}\label{delB}
\Gamma^a{}^{C(A}\,\delta\cals B_{a}{}^{B)}{}_C\eql 0\,,
\end{equation}
\begin{equation}\label{delA}
3\,\Gamma^{a\,[AB}\,\delta\cals B_a{}^{C]}{}_D-3 \Gamma^a_{EF}\,\delta\cals A_{a}^{EF[AB}\delta^{C]}{}_D-4 \,\delta\cals A_a{}^{ABCE}\Gamma^a_{ED}\eql 0\,.
\end{equation}
Those are equations of the type we have already encountered and solved
above in \eqref{homvieleqs}, \eqref{A1eqsor} and \eqref{A2eqss}, so we may readily write the solution
\begin{equation}\label{delevAA1}
e_a{}^m\delta \cals B_m{}^A{}_B   \eql -4 \,i\,\delta X_{a|b}\,\Gamma^b_{AB}-\coeff2 3\,\delta X_{a|bcd}\,\Gamma^{bcd}_{AB}\,,
\end{equation}
\begin{equation}\label{delevAA2}
\begin{split}
e_a{}^m\delta\cA_m{}_{ABCD} & \eql i\,\delta X_{a|b}\,\Gamma^c_{[AB}\Gamma^{ac}_{CD]}+
 \delta X_{a|bcd}\,\Gamma^{b}_{[AB}\Gamma^{cd}_{CD]}\,,
\end{split}
\end{equation}
where 
\begin{equation}\label{thecorrection}
\delta X_{a|a}\eql 0\,,\qquad \delta X_{[a| bcd]}\eql 0\,.
\end{equation}
While it is straightforward to verify the solution for \eqref{delB}, the proof for \eqref{delA} is more involved. We proceed similarly as before by first substituting \eqref{delevAA1} and \eqref{delevAA2} into
\eqref{delA} and then contracting  this equation with $\Gamma^a_{AB}$ 
and $\Gamma^{ab}_{AB}$, respectively,  to show that the resulting expressions
indeed vanish  if \eqref{thecorrection} is satisfied. Again we omit the lengthy 
intermediate expressions.

This proves that given any solution on the $d\!=\!4$ side, we can always correct it such as to obtain after the $U$-rotation a solution with the tensor structure consistent with the $d\!=\!11$ supersymmetry. Moreover, that solution is unique and hence determines the fluxes $f$ and $F_{abcd}$ in terms of the scalar vielbein of the $\cN\!=\!8$, $d\!=\!4$ theory. Furthermore, the constraints \eqref{thecorrection} on the homogeneous correction are precisely such that both $f$ and $F_{abcd}$ do not get modified in the process, and hence, at a given point on the scalar coset, can be read off from {\it any} solution on the  $d\!=\!4$ side, even if that solution may not be the one that has the correct tensor structure after it is $U$-rotated to $d\!=\!11$.
This both completes the proof of the consistent truncation in \cite{deWit:1986iy} and shows that one can extract correct fluxes from the standard solution that was found there.

\section{The fluxes}
\label{secfive}
\setcounter{equation}{0}

In this section, we will outline in a systematic way  two methods  for computing the  fluxes, $f$ and $F_{mnpq}$,  in terms of $d\!=\!4$ quantities. The first method follows directly from the discussion above  and has been essentially spelled out already. The second one requires some additional work to prove the   flux formulae \eqref{thefproj} and \eqref{theFproj}. 

For  a fixed scalar 56-bein \eqref{V} in $d\!=\!4$, the starting point for computing the corresponding field configuration in $d\!=\!11$ is the triplet,
\begin{equation}\label{}
e^m_{\,ij}\,,\qquad \cals B_m{}^i{}_j\,,\qquad \cals A_{m\,ijkl}\,,
\end{equation}
where $e^m_{\,ij}$ is the generalized vielbein defined in \eqref{genvieldef} and 
$(\cals B_m{}^i{}_j\,,\cals A_{m\,ijkl})$ is a solution to the {\it GVP\/} \eqref{vieleqs} 
and the $\frak A$-equations \eqref{frakAeqs1} and \eqref{frakAeqs2}. One can either use the standard inhomogeneous solution \eqref{theBzero}-\eqref{specalpha}, or any other solution if that is more convenient. It follows from the general solution to the {\it GVP\/} in Section \ref{sectwo} that, for a given generalized vielbein, $\cals B_m{}^i{}_j$ and $\cals A_{m\,ijkl}$ completely determine each other. Hence either of them contains the full information about the fluxes. Here 
we  choose to work with   $\cals A_{m\,ijkl}$ for the simple reason that  it is an $\rm SU(8)$ tensor.

From the generalized vielbein we  determine the metric, $g_{mn}$, and the warp factor, $\Delta$, which in general are already quite difficult to obtain in a closed analytic form.

Next we turn to the fluxes. In the first method, we calculate the metric  vielbein, $e_m{}^a$, and  solve  \eqref{aligngv} for the $\rm SU(8)$ rotation matrix, $U$, and then use the latter to rotate $\cals A_{m\,ijkl}$ to $d\!=\!11$ according to  \eqref{Utransform}. The resulting $\cals A_{m\,ABCD}$ tensor is a solution to the {\it GVP} \eqref{vieleqs11} and thus of the general form given in \eqref{Ad11}. We then read off the expansion coefficients, $X_{a|b}$ and $X_{a|bcd}$, from which the fluxes are obtained by projecting onto   irreducible $\rm SO(7)$ components, see \eqref{specfsol},
\begin{equation}\label{projfl}
f\eql 4\sqrt 2\,X_{a|a}\,,\qquad F_{abcd}\eql 16\sqrt 2\,X_{[a|bcd]}\,.
\end{equation}
As was shown above, the result does {\em not} depend on which particular solution 
$\cals A_{m\,ijkl}$ of the {\it GVP} we start  with on the $d\!=\!4$ side.

A difficulty one encounters in trying to apply this construction in any specific example is that  both the metric vielbein, $e_m{}^a$,  and the $\rm SU(8)$ rotation, $U$, are obtained by solving quadratic equations, and can become quite complicated, if calculable analytically at all. Hence,    one would like to avoid having to 
perform the rotation explicitly by working with $\rm SU(8)$ invariant quantities as 
in \eqref{metrivfor}. The method how to do this was outlined in \cite{deWit:1986iy}, 
and here we  expand on it.

We start on the $d\!=\!11$ side with  $\cals A_{m\,ABCD}$  that is obtained from $\cals A_{m\,ijkl}$ by the $U$-rotation \eqref{Utransform}. Then the fluxes are given by the projections \eqref{projfl} of the expansion coefficients of $\cals A_{m\,ABCD}$ in \eqref{Ad11}, so all that is needed is an effective way to extract those two projections from the $d\!=\!4$ result. The problem here is that the basis of the $\Gamma$-matrices used in \eqref{Ad11} is not $\rm SU(8)$-covariant and $U$-rotation mixes different terms in the expansion. 

In order to construct $\rm SU(8)$-covariant projections, we note  that \eqref{genvield11} and \eqref{aligngv} imply a covariant transformation between the $\Gamma$-matrices on the $d\!=\!11$ side and the generalized vielbeine on the $d\!=\!4$ side, namely
\begin{equation}\label{Uonviel}
\Gamma^a_{AB} U^A{}_iU^B{}_j\eql -i\,\Delta^{1/2}\,e_m{}^a\,e^m_{\,ij}\,,
\end{equation}
and
\begin{equation}\label{Uonvielstar}
\Gamma^a_{AB} (U^A{}_i)^*(U^B{}_j)^*\eql -i\,\Delta^{1/2}\,e_m{}^a\, e^{m\,ij}\,,
\end{equation}
and it is this covariance that must be preserved. In particular, it implies that
an  $\rm SU(8)$-covariant basis for the expansion of the 
$\cals A_{m\,ABCD}$ tensor must be constructed from odd products of $\Gamma$-matrices, which are then  
$U$-rotated into $\rm SU(8)$-invariant contractions of the vielbeine 
$e^m_{\,ij}$ and $e^{m\,ij}$.
To implement this change of basis, we use $\Gamma$-matrix identities,
\begin{equation}\label{twoGafive}
\begin{split}
\Gamma^{ab}_{AB} & \eql -\coeff i {5!}\,\epsilon^{abcdefg}\Gamma^{cdefg}_{AB}\,,\qquad 
\Gamma^{abcde}_{AB}   \eql \coeff i 2\, \epsilon_{abcdefg}\,\Gamma^{fg}_{AB}\,,
\end{split}
\end{equation}
and
\begin{equation}\label{gafiveid}
\Gamma^a_{[AB}\Gamma^{bcdef}_{CD]}\eql \Gamma^{[a}_{[AB}\Gamma^{bcdef]}_{CD]}
+\coeff 5 3\,\delta^{a[b}\Gamma^{|g|}_{[AB}\Gamma^{cdef]g}_{CD]}\,,
\end{equation}
to recast \eqref{Ad11} in the form
\begin{equation}\label{Acovexp}
\cals A_{m\,ABCD}\eql -\coeff34 \, (S^{-1} \Do_mS)_{ab}\Gamma^{a}_{[AB}\Gamma^b_{CD]}+i\,
X_{m\,abcd}\,\Gamma^f_{[AB}\Gamma^{abcdf}_{CD]}+
X_{m\,abcdef}\,\Gamma^{[a}_{[AB}\Gamma^{bcdef]}_{CD]}\,,
\end{equation}
where $X_{m\,abcd}$ and $X_{m\,abcdef}$ are completely antisymmetric in 
their (flat) indices, and are related to the original expansion coefficients in \eqref{Ad11} by 
\begin{equation}\label{xfour}
X_{m\,abcd}\eql -\coeff 1{3\cdot 4!}\epsilon_{abcdefg}\,X_{m\,efg} \,,
\end{equation}
\begin{equation}\label{xsix}
\begin{split}
X_{m\,abcdef}\eql \coeff 1{5!} \epsilon_{abcdefg}\,X_{m\,g} \,.
\end{split}
\end{equation}
Then from \eqref{projfl} we obtain\footnote{The indices on the Levi-Civita symbol  are raised and lowered with the background metric.}
\begin{equation}\label{fformula}
f\eql 4\sqrt 2\,e_a{}^mX_{ma}\eql \coeff{2\sqrt 2}{3}\,\epsilon^{abcdefg}\,e_a{}^m\,X_{m\,bcdefg}\,,
\end{equation}
and
\begin{equation}\label{thefourF}
F_{abcd}\eql -8\sqrt2\,e_{[a}{}^m\epsilon_{bcd]}{}^{efgh}X_{m\,efgh}\,.
\end{equation}
One should note the  antisymmetrization in \eqref{thefourF} that projects onto the correct   tensor structure of the  four-form flux in $d\!=\!11$ supergravity.

To project out the components \eqref{xfour}
 and \eqref{xsix} from the $\cals A_{m\,ABCD}$ tensor, we can simply contract with the basis tensors $\Gamma^f_{[AB}\Gamma^{abcdf}_{CD]}$ and $ \Gamma^{[a}_{[AB}\Gamma^{bcdef]}_{CD]}$, which are orthogonal with respect to each other and normalized according to 
\begin{equation}\label{}
\Gamma^{c}_{[AB}\Gamma^{a_1a_2a_3a_4c}_{CD]}\,\Gamma^{d}_{AB}\Gamma^{b_1b_2b_3b_4d}_{CD}\eql 
48\cdot 4!\,\delta^{a_1a_2a_3a_4}_{b_1b_2b_3b_4}\,,
\end{equation}
\begin{equation}\label{}
\Gamma^{[a_1}_{[AB}\Gamma^{a_2a_3a_4a_5a_6]}_{CD]}\Gamma^{[b_1}_{AB}\Gamma^{b_2b_3b_4b_5b_6]}_{CD}\eql 32\cdot 5!\,\delta^{a_1a_2a_3a_4a_5a_6}_{b_1b_2b_3b_4b_5b_6}\,.
\end{equation} 
The same projections in terms of $\rm SU(8)$-covariant products  of the vielbeine, 
$e^m_{ij}$ and $e^{m\,ij}$,  obtained using \eqref{Uonviel} and \eqref{Uonvielstar}, can 
be applied to $\cals A_{m\,ijkl}$. In this way, after passing to curved indices, we obtain explicit formulae for the fluxes expressed entirely in terms of $d\!=\!4$ quantities:
\begin{equation}\label{theffinal}
f\eql -\coeff {\sqrt 2}{48\cdot 5!}\,\Delta^4\,g^{mu}\,\epsilon_{mnpqrst}\,e^n_{ij}\,(e^{[p}\bar e^q e^r\bar e^s e^{t]})_{kl}\,\cals A_u{}^{ijkl}\,,
\end{equation}
and
\begin{equation}\label{theFfinal}
F_{mnpq}\eql -\coeff{i}{144}\,\Delta^4\,g_{rw}\,e^r_{ij}(e^{[s}\be^te^u\be^ve^{w]})_{kl}\,\epsilon_{stuv[mnp}\,\cals A_{q]}{}^{ijkl}\,.
\end{equation}
Those expressions for the fluxes are valid for any $\cals A_{m\,ijkl}$ satisfying the {\it GVP\/} 
and the $\frak A$-equations. In particular, they hold for $\cAo_{m\,ijkl}$, the standard solution \eqref{theAzero}.

For the particular $\cals A_{m\,ijkl}$ that   has the correct tensor structure in $d\!=\!11$, we may combine 
\eqref{theffinal} and \eqref{theFfinal} into \eqref{fluxeqs}, however, for other  $\cals A_{m\,ijkl}$ this equation will in general not hold if there are other than just the flux components in $\cals A_{m\,ABCD}$. We will illustrate this on some examples in the next section. 

Finally, let us note that by plugging in the general solution for $\cals A_{m\,ijkl}$ derived in Sections \ref{sectwo} and \ref{secthree}, and parametrized in terms of $\alpha_{a|b}$ and $\alpha_{a|bcd}$, in the flux formulae \eqref{theffinal} and \eqref{theFfinal}, we can directly relate two of the $X$-parameters, namely, $X$ and $X_{abcd}^{\,\Yboxdim2.5pt\yng(1,1,1,1)}$, to the $\alpha$-parameters. Working out projection  formulae for the other $\rm SO(7)$ irreducible components, similar relations can be obtained for other parameters as well. However, it is clear that, largely  because of the $\rm E_{7(7)}$ rotation that mixes different $\Gamma$-matrix structures, the final formulae will be quite involved and not very illuminating.

\section{Analytic examples}
\label{secsix}
\setcounter{equation}{0}

In this section we illustrate various points of the general discussion on three examples: (i) the $\rm SO(8)$ critical point, (ii) the $\rm SO(7)^-$ invariant family, and (iii) the $\rm SO(7)^+$ invariant family, for which  the $U$-rotations are known in a closed form.

\subsection{$\rm SO(8)$}

We begin with the maximally supersymmetric critical point,
\begin{equation}\label{}
u_{ij}{}^{IJ}\eql\delta_{ij}{}^{IJ}\,,\qquad v_{ijIJ}\eql 0\,,
\end{equation}
corresponding to the $AdS_4\times S^7$ solution \cite{EFI 80/35-CHICAGO} of $d\!=\!11$ supergravity, 
\begin{equation}\label{FrRsol}
e_m{}^a=\eo_m{}^a\,,\qquad f   \eql 3\sqrt 2\,m_7\,,\qquad F_{abcd} \eql0\,.
\end{equation}

The generalized vielbein and the $U$-rotation are simply, \begin{equation}\label{}
e^m_{\,ij}\eql i\,\eo_a{}^m\,\bar\eta^i\Gamma^a\eta^j\,,\qquad U^i{}_A\eql\eta^i{}_A\,,
\end{equation}
and the general solution to the {\it GVP\/} and the $\frak A$-equations, c.f.\ \eqref{thed4sol},  is
\begin{equation}\label{Atrivial}
\eo_a{}^m\,A_{m\,ijkl}\eql -\coeff i 4 \,(-\coeff 3 7\,m_7 \,\delta_{ab}+\alpha^{\,\Yboxdim4pt\yng(2)}_{ab}+\alpha^{\,\Yboxdim4pt\yng(1,1)}_{ab})\,\bar\eta_{[i}\Gamma^{bc}\eta_j\bar\eta_k\Gamma^c\eta_{l]}
-\coeff 3 2\,(\alpha^{\,\Yboxdim4pt\yng(2,1,1)}_{abcd}+
\delta^{\phantom{{\,\Yboxdim4pt\yng(1,1)}}}_{a[b}\tilde\alpha_{cd]}^{\,\Yboxdim4pt\yng(1,1)})\,
\bar\eta_{[i}\Gamma^{[bc}\eta_j\bar\eta_k\Gamma^{d]}\eta_{l]}\,.
\end{equation}
Since the $U$-rotation is merely a change between the two types of $\rm SU(8)$ indices,
the $X$-parameters are proportional to the  $\alpha$-parameters,
\begin{equation}\label{}
 X_{a|b}\eql -\coeff 1 4\,\alpha_{a|b}\,,\qquad \alpha_{a|bcd}\eql -\coeff 2 3\, X_{a|bcd}\,,
\end{equation}
and, in particular, we have
\begin{equation}\label{}
X\eql \coeff 3 4\,m_7\,,\qquad  X_{abcd}^{\,\Yboxdim4pt\yng(1,1,1,1)}\eql 0\,,
\end{equation}
from which the fluxes \eqref{FrRsol} follow. 

We also note that by setting the explicit $\alpha$-parameters in \eqref{Atrivial} to zero, we recover  the standard inhomogenous solution, which in this example satisfies the correct tensor structure condition. This is not surprising, as by  construction the standard inhomogenous solution \eqref{theBzero} and \eqref{theAzero} has the same symmetry as the scalar background in $d\!=\!4$, and the only $\rm SO(8)$-invariant $\alpha$-parameter that one can have is a constant singlet.  The formula for the fluxes  based on the standard solution  was already tested in   \cite{deWit:1986iy}.

We may also verify  the fluxes   by evaluating the   projections \eqref{theffinal} and \eqref{theFfinal}.
From the  orthonormality of the Killing spinors and  $\Gamma$-matrix identities, we have
\begin{equation}\label{controne}
\epsilon_{mnpqrst}\, e^n_{ij}\,(e^{[p}\bar e^q e^r\bar e^s e^{t]})_{kl}\,\bar\eta_{[i}\Gamma^{bc}\eta_j\bar\eta_k\Gamma^c\eta_{l]}  \eql 192\cdot 5!\,i\,\eo_m{}^a\,,
\end{equation}
and
\begin{equation}\label{contrtwo}
\epsilon_{mnpqrst}\, e^n_{ij}\,(e^{[p}\bar e^q e^r\bar e^s e^{t]})_{kl}\,\bar\eta_{[i}\Gamma^{[bc}\eta_j\bar\eta_k\Gamma^{d]}\eta_{l]}   \eql 0\,,
\end{equation}
where  \eqref{contrtwo} also follows from the $\rm SO(8)$ invariance.
 Substituting those contractions in 
\eqref{theffinal} and using the tracelessness of    $\alpha^{\,\Yboxdim4pt\yng(2)}_{ab}$ and $\alpha^{\,\Yboxdim4pt\yng(1,1)}_{ab}$, we get
\begin{equation}\label{}
f\eql -(\coeff {\sqrt 2}{48\cdot 5!})\times (\coeff 3{28}i)\times (192\cdot 5!\,i)\times 7\,m_7\eql 3\sqrt 2\,m_7\,.
\end{equation}
The vanishing of the internal flux is verified similarly.

\subsection{$\rm SO(7)^-$}

The $\rm SO(7)^-$-invariant sector  of $\cN\!=\!8$, $d\!=\!4$ supergravity provides  the simplest example with a nontrivial $U$-rotation \cite{deWit:1983vq}. By symmetry, the standard inhomogenous solution must have the correct tensor structure in $d\!=\!11$ and we verify 
that at the $\rm SO(7)^-$ critical point  it  yields the  fluxes of Englert's  `parallelizing torsion 
solution' of $\cN\!=\!1$, $d\!=\!11$ supergravity  
\cite{CERN-TH-3394}. 

The scalar 56-bein in this sector forms  a one parameter family \cite{deWit:1983gs}
\begin{equation}\label{so7m56}
u_{ij}{}^{IJ} (t) \eql u_1(t)\,\delta_{ij}^{IJ}+u_2(t)\,C^{ijIJ}_-\,,\qquad
v_{ij}{}_{IJ} (t) \eql v_1(t)\,\delta_{ij}^{IJ}+v_2(t)\,C^{ijIJ}_-\,,
\end{equation}
where $C^{IJKL}_-$ is an anti-selfdual tensor satisfying
\begin{equation}\label{}
C^{IJMN}_-C^{MNKL}_-\eql 12\delta^{IJ}_{KL}-4C_-^{IJKL}\,, 
\end{equation}
and
\begin{equation}
\begin{aligned}
u_1(t) & \eql \cosh^3(2t)\,,\qquad &u_2(t) \eql \coeff 1 2 \cosh(2t)\sinh^2(2t)\,,\\ 
v_1(t) & \eql i \sinh^3(2t)\,,\qquad 
&v_2(t) \eql \coeff i 2 \cosh^2(2t)\sinh(2t)\,.
\end{aligned}
\end{equation}
The scalar potential along the $\rm SO(7)^-$ family is
\begin{equation}\label{Pinso7m}
\cals P (t) \eql  -2\,g^2 \cosh^{5}(4t) (5-2\cosh(8t))\,,
\end{equation}
and has two critical points: the maximally supersymmetric one at $t=0$, and the $\rm SO(7)^-$ point at 
 $t=\coeff 1 4\,{\rm arcoth}(\sqrt5)$. 

The  solution of $d\!=\!11$ supergravity  corresponding to the $\rm SO(7)^-$ point can be  expressed entirely in terms of an $\rm SO(7)^-$ invariant rank three tensor, $S_{abc}$, on $S^7$,
\begin{equation}\label{paraltor}
S_{abc}\eql {i\over 16}\,C_-^{IJKL}\,\bar\eta^I \Gamma_{[ab}\eta^J\,\bar\eta^K\Gamma_{c]}\eta^L\,,
\end{equation}
known as the `parallelizing torsion,' in terms of which the generalized vielbein is given by
\begin{equation}\label{so7mviel}
e^m_{\,ij}\eql i\,(u_1+v_1)\,\bar\eta^i\Go^m\eta^j+(u_2+v_2)\,S_{mab}\,\bar\eta^i\Gamma^{ab}\eta^j\,.
\end{equation}
As usual, the conversion between the flat/curved indices on the $\Gamma$-matrices and the $S_{abc}$ tensor is  done with the background vielbeins.
To derive \eqref{so7mviel}, one uses the `inverse' of \eqref{paraltor}, see \cite{deWit:1983vq},
\begin{equation}\label{StoC}
C_-^{IJKL}\eql \coeff i 2\,S_{abc}\,\bar\eta^{[I}\Gamma^{ab}\eta^J\bar\eta^K\Gamma^c\eta^{L]}\,.
\end{equation}
Then, using\footnote{For a  full list of identities satisfied by torsion tensor, see \cite{deWit:1983gs}. They imply that any contraction and background derivative of the torsion tensor(s) 
can be reduced to terms that are linear in it. } 
\begin{equation}\label{}
S_{acd}S_{bcd}\eql 6\,\delta_{ab}\,,
\end{equation}
the metric and the warp factor are calculated from the metric lift formula \eqref{metrivfor}:
\begin{equation}\label{Delso7m}
\Delta^{-1} g^{mn}\eql \cosh^3(4t)\,\go^{mn}\,,\qquad \Delta^{-1}=\cosh^{7/3}(4t)\,.
\end{equation}
The metric  vielbein, $e_m{}^a$,  can be chosen to be proportional to $\eo_m{}^a$,
\begin{equation}\label{myvielbein}
e_m{}^a\eql \cosh^{-1/3}(4t)\eo_m{}^a\,,
\end{equation}
which means that the $S$-terms in $\cals A_{m\,ABCD}$ and $\cals B_m{}^A{}_B$ 
will be absent. As we already pointed out, once the normalization at the SO(8) point is fixed,
the overall normalization of the $d\!=\!11$ metric (as well as the other fields) is fixed 
along the whole family, and in particular at the SO(7)$^-$ stationary point.

The $U=\Phi\eta$ matrix for those vielbeine was calculated in \cite{deWit:1983vq}, 
\begin{equation}\label{}
\Phi
\eql \coeff 1 8(e^{-7i\tau}+7e^{i\tau})+\coeff i {48}(e^{-7i\tau}-e^{i\tau})S_{abc}\Gamma^{abc}\,,
\end{equation}
where the parameter $\tau$ is related to $t$ by
\begin{equation}\label{ttotau}
\tan(2\tau)\eql\tanh(2t)\,.
\end{equation}

To evaluate $\cAo_{m\,ijkl}$ of the standard inhomogenous solution \eqref{theAzero}, 
we first rewrite the Killing vectors  in terms of the Killing spinors \eqref{Kvectdef}. It follows from \eqref{Kvectdef} and \eqref{kspeqs} that 
\begin{equation}\label{derKillVect}
\Do_mK_n{}^{IJ}\eql -m_7\,\bar\eta^I\Go_{mn}\eta^J\,,
\end{equation}
and we use it to similarly rewrite the second term in \eqref{theAzero}. Finally, after using   \eqref{StoC} in the scalar 56-bein \eqref{so7m56}, and the orthonormality of the Kiling spinors, we are left with some tedious $\Gamma$-matrix algebra that is required to simplify the resulting expressions.  A number of useful identities to do that can be found in \cite{LPTENS 79/6} and \cite{deWit:1986mz}. The result is
\begin{equation}\label{Aso7md4}
\cAo_{m\,ijkl}\eql -\coeff{i}{28}\,m_7\sinh(8t) \eo_{md}\,S_{abc}\eta^{[i}\Gamma^{da}\eta^j\bar\eta^k\Gamma^{bc}\eta^{l]}+\coeff {3 \,i}{28}\, m_7\cosh(4t) \eo_{mb}\bar\eta_{[i}\Gamma^{ba}\eta_j\bar\eta_{k}\Gamma^a\eta_{l]}\,.
\end{equation}

The $U$-rotation  to $d\!=\!11$ requires similar algebra, but is even more tedious  and 
we omit here the intermediate steps.  At the end we find
\begin{equation}\label{rotAtens}
\begin{split}
\cAo_{m\,ABCD} & \eql \coeff i{28} \,m_7 (5-2\cosh(8t))\eo_m{}^a\Gamma^{ab}_{[AB}\Gamma^b_{CD]}\\
& \hspace{1 in} +\coeff 1 {48} \,m_7 \sinh(4t)\eo_m{}^a\epsilon_{abcdefg}S_{efg}\Gamma^{b}_{[AB}\Gamma^{cd}_{CD]}\,.
\end{split}
\end{equation}
We see that since $e_m{}^a$ and $\eo_m{}^a$ are proportional, this tensor has the correct tensor structure \eqref{su8ABA} and we readily read off the fluxes not just at the critical points, but  along the entire $\rm SO(7)^-$ family,
\begin{equation}\label{ffromlift}
f\eql \sqrt 2 m_7 \cosh^{1/3}(4t) (5-2\cosh(8t))\,,
\end{equation}
\begin{equation}\label{Ffromlift}
F_{abcd}\eql \coeff {\sqrt 2}{3} \,m_7\cosh^{1/3}(4t) \sinh(4t)\,\epsilon_{abcdefg}S_{efg}\,.
\end{equation}
Once more the correct tensor structure of \eqref{rotAtens} is guaranteed by the $SO(7)^-$ symmetry -- one cannot construct from the torsion tensor, $S_{abc}$, and the vielbein, $e_m{}^a$,  any other coefficients $X_{a|b}$ and $X_{a|bcd}$  than the singlet and the completely antisymmetric tensor, respectively. However, the normalization of each term at the $\rm SO(7)^-$ critical point must agree with the known solution, which is a nontrivial test of our flux formulae. 

\def\wt{\widetilde}

The solution  in \cite{deWit:1983gs,deWit:1983vq} is\footnote{Since the same symbol in \cite{deWit:1983gs,deWit:1983vq} may denote quantities that are related by a rescaling to the ones here, we put a tilde whenever the identification is not immediately obvious.  }
\begin{equation}\label{}
\wt f\equiv -\coeff i{24}\,\Delta^{-1/2}\epsilon^{\alpha\beta\gamma\delta}F_{\alpha\beta\gamma\delta}\eql\sqrt 2\, \wt m_7\, (3-4\tan^2(4\tau))\,,
\end{equation}
\begin{equation}\label{}
F_{abc}\eql \coeff 1 {24}\, \Delta^{-1/2}\epsilon_{abc}{}^{defg}F_{defg}\eql 2\sqrt 2\,\wt m_7\,\tan(4\tau)S_{abc}\,,
\end{equation}
where the flux components, $F_{\alpha\beta\gamma\delta}$ and $F_{abcd}$, are with respect to the vielbein of the same form as in \eqref{myvielbein}, while $\wt f$ is rescaled with respect to \eqref{fluxsol},
\begin{equation}\label{}
\wt f\eql\Delta^{-1/2}\, f\,.
\end{equation}
Then, noticing that from \eqref{ttotau} we have
\begin{equation} 
5-2\cosh(8t) = 3 - 4\tan^2 (4\tau) \;, \quad \sinh(4t) = \tan (4\tau)   \;,
\end{equation}
we get
\begin{equation}\label{fsol}
f
\eql\sqrt 2\,\wt m_7\,\Delta^{1/2}\,(5-2\cosh(8t))\,,
\end{equation}
\begin{equation}\label{}
\begin{split}
F_{abcd}  \eql \coeff 1 6  \Delta^{1/2}\epsilon_{abcd}{}^{efg}F_{efg}
  \eql \coeff{\sqrt 2}3\,  \wt m_7\,  \Delta^{1/2}\,\sinh(4t) \epsilon_{abcdefg}S_{efg}\,.
\end{split}
\end{equation}
We see that those fluxes agree with the ones in \eqref{ffromlift} and \eqref{Ffromlift}, provided we set
\begin{equation}\label{tilmseven}
\wt m_7\eql \cosh^{3/2}(4t)\,m_7\,.
\end{equation}
There is no explicit expression for $\wt m_7$ as a functions of $t$ in \cite{deWit:1983vq}, except that its relation to the gauge coupling constant, $g$, changes along the family. Since we have $g=\sqrt 2\,m_7$, we deduce from \eqref{tilmseven} that this relation must be
\begin{equation}\label{}
g\eql \sqrt 2\,\cosh^{-3/2}(4t)\,\wt m_7\,.
\end{equation}
This implies that at the maximally supersymmetric point,
\begin{equation}\label{}
g\eql \sqrt 2\,\wt m_7\,,
\end{equation}
while at the $\rm SO(7)^-$ point,
\begin{equation}\label{}
g\eql 4\cdot 5^{-3/4}\,\wt m_7\,.
\end{equation}
This agrees with (1.4) in \cite{deWit:1983vq} and is our first new non-trivial test of the lift formulae for the fluxes.

\subsection{$\rm SO(7)^+$}

The $\rm SO(7)^+$ solution of $\cN\!=\!1$, $d\!=\!11$ supergravity is constructed in  terms of an invariant vector field $\xi^a$ on $S^7$ \cite{deWit:1984va}. This implies that  the $X_{a|bcd}$ parameters of the standard inhomogenous solution in $d\!=\!11$ must vanish, 
which agrees with $F_{abcd}=0$ \cite{deWit:1984va}, but allows for  $X_{a|b}$ 
with both   trace and   traceless components. We will show that in fact the standard 
inhomogeneous solution  has a non-vanishing $X^{\,\Yboxdim4pt\yng(2)}_{ab}$ term, which can be removed by a suitable homogeneous correction, and that the resulting flux, $f$, 
agrees with the known solution. Throughout this subsecton we set $m_7=1$.

The scalar 56-bein of the one-parameter $\rm SO(7)^+$ invariant family is
\begin{equation}\label{so7pv}
u_{ij}{}^{IJ}(t)\eql u_1(t)\,\delta^{IJ}_{ij}+u_2(t)\,C^{ijIJ}_+\,,\qquad 
v_{ij}{}_{IJ}(t)\eql v_1(t)\,\delta^{IJ}_{ij}+v_2(t)\,C^{ijIJ}_+\,,
\end{equation}
where $C^{IJKL}_+$ is a self-dual tensor satisfying
\begin{equation}\label{}
C^{IJMN}_+C^{MNKL}_+\eql 12\,\delta^{IJ}_{KL}+4\,C_+^{IJKL}\,, 
\end{equation}
and
\begin{equation}\label{}
\begin{aligned}
u_1(t)& \eql \cosh^3(2t)\,,\qquad & u_2(t)\eql \coeff 1 2 \cosh(2t)\sinh^2(2t)\,,\\
v_1(t)& \eql \sinh^3(2t)\,,\qquad & v_2(t) \eql \coeff 1 2 \cosh^2(2t)\sinh(2t)\,.
\end{aligned}
\end{equation}
All the dependence on the internal geometry can be expressed in terms of the vector field,
\begin{equation}\label{defofxi}
\xi^a(y)\eql \coeff i {16}\,C_+^{IJKL}\,\bar\eta^I\Gamma^{ab}\eta^J\,\bar\eta^K\Gamma^b\eta^L\,,
\end{equation}
and a scalar function, $\xi(y)$, defined by
\begin{equation}\label{}
\xi^a\xi^a\eql (3-\xi)(21+\xi)\,.
\end{equation}
 A complete list of identities satisfied by $\xi^a$ and $\xi$ and their derivatives as well as  further discussion of their properties can be found in  \cite{deWit:1984va}.
 
Once more, \eqref{defofxi} can be inverted by \cite{deWit:1984va},
\begin{equation}\label{}
\begin{split}
C_+^{IJKL}\eql \coeff 1 {12}\,(9+\xi)\, & \bar\eta^{[I}\Gamma^a\eta^J\bar\eta^K\Gamma^a\eta^{L]}\\& -{\xi^a\xi^b\over 4(3-\xi)}\,\bar\eta^{[I}\Gamma^a\eta^J\bar\eta^K\Gamma^a\eta^{L]} +\coeff i {12}\,\xi^a\,\bar\eta^{[I}\Gamma^{ab}\eta^J\bar\eta^K\Gamma^b\eta^{L]}\,,
\end{split}
\end{equation}
which is then used to rewrite the 56-bein \eqref{so7pv}. It is then straightforward to obtain
the generalized vielbein \eqref{genvieldef}, which reads
\begin{equation}\label{genvielso7p}
\begin{split}
e^m_{\,ij}
& \eql  \left[(u_1+v_1)-\coeff 1 3(u_2+v_2) (3+\xi)\right]\,i\, \bar\eta^i\Gamma^m\eta^j\\
& \hspace{1 in} + (u_2+v_2)\,{\xi^m\,\xi^a\,\over 3( 3-\xi)} \,i\,\bar\eta^i\Gamma^a\eta^j
+\coeff 1 3 (u_2+v_2)\,\xi^a\,\bar\eta^i\Gamma^{am}\eta^j\,,
\end{split}
\end{equation}
where the conversion to curved indices is with the background vielbein, $\eo_a{}^m$. 

At this point it is convenient to switch to the parameter,
\begin{equation}\label{}
\tau\eql \frac{e^{8 t}-1}{3 \left(e^{8 t}+7\right)}\,,
\end{equation}
introduced in \cite{deWit:1984va},
and absorb any dependence on $\xi$ into the function
\begin{equation}\label{}
H(\xi,\tau)\eql {(1+21\tau)\over{\sqrt{(1+63\tau^2)-2\,\xi\,\tau\,(1+9\tau)}}}\,.
\end{equation}
In the new parametrization, the maximally supersymmetric critical point is at $\tau=0$, while the $\rm SO(7)^+$ critical point is at $\tau\eql \coeff1 {33}\,(2\sqrt 5-3)$ or, equivalently,
at $16\,t = \ln 5$.

From the generalized vielbein \eqref{genvielso7p}, one calculates the metric, the warp factor and the metric vielbein.  The latter is given by \cite{deWit:1984nz}
\begin{equation}\label{so7pviel}
e_m{}^a\eql \lambda^{1/2}\left[\delta_m{}^a-\left(1-{1\over H}\right)\,{\bxi_m\bxi^a}\right]\,,
\end{equation}
where
$\bxi^a$ is the unit vector field corresponding to $\xi^a$, and 
\begin{equation}\label{thedeff}
 \lambda\eql {(1-3\tau)^{1/3}\over (1+21 \tau)^{1/3}}\,H^{2/3}\,.
\end{equation}
The warp factor  is
\begin{equation}\label{so7pwarp}
\Delta\eql  {(1-3\tau)^{7/6}\over (1+21 \tau)^{7/6}}\,H^{4/3}\,.
\end{equation}

We note that the overall $\tau$-dependent normalization factors in \eqref{thedeff} and \eqref{so7pwarp}, that follow  from the lift formula for the metric \eqref{metrivfor}, differ from those 
in \cite{deWit:1984nz}  which were obtained by a different method. While this is irrelevant 
for a solution at the critical point, the correct $\tau$-dependent  normalization is needed 
to obtain  lifts of more general solutions, such as RG-flows.

The $U=\Phi\eta$ matrix was given in \cite{deWit:1984nz} and it reads
\begin{equation}\label{}
\Phi\eql \cos\vartheta +\sin\vartheta \,(i\,\bxi^a\Gamma^a)\,,
\end{equation}
where
\begin{equation}\label{}
\cos(2\vartheta)\eql {H\over (1+21\tau)}(1-\tau\xi)\,,\qquad
\sin(2\vartheta)\eql  {H\tau\over (1+21\tau)}\sqrt{(21+\xi)(3-\xi)}\,.
\end{equation}
Note that in this example  $\Phi$ (and $U$) is a real and hence an orthogonal matrix.

The evaluation of the   $\cAo_{m\,ijkl}$ tensor \eqref{theAzero} of the standard inhomogenous solution follows the same steps as in the $\rm SO(7)^-$ example, but is much more tedious as may be inferred from the presence of higher rank symmetric tensors that can be constructed from products of $\xi^a$'s. We refer the reader to \cite{deWit:1986mz,deWit:1984va} for some useful identities.

Let us expand the tensor  $\cals A_{m\,ijkl}(\alpha,\beta)$  on the left hand side in \eqref{theAzero}  as 
\begin{equation}\label{}
\cA_{m\,ijkl}(\alpha,\beta)\eql \alpha \,\cals A_{m\,ijkl}^{(\alpha)}+\beta\,\cals A_{m\,ijkl}^{(\beta)}\,.
\end{equation}
The tensors on the right hand side,  written in the form similar to \eqref{genvielso7p} that allows us to trace the origin of individual terms,  are given by 
\begin{equation}\label{Aalso7}
\begin{split}
A^{(\alpha)}_{m\,ijkl} & \eql (u_1v_2-u_2v_1)\,\Big[\,\coeff 1 2 \,\xi^a\,\bar\eta_{[i}\Gamma_m\eta_j\bar\eta_k\Gamma^a\eta_{l]}-\coeff 1 6 \,\xi_m\,\bar\eta_{[i}\Gamma^a\eta_j\bar\eta_k\Gamma^a\eta_{l]}\\[4 pt]
& \hspace{1.25in} -\coeff i {12} \,(3-\xi)\,\bar\eta_{[i}\Gamma_m{}^a\eta_j\bar\eta_k\Gamma^a\eta_{l]}+\coeff i {12} \,{\xi_m\xi^a\over (3-\xi)} \,\bar\eta_{[i}\Gamma^{ab}\eta_j\bar\eta_k\Gamma^b\eta_{l]}\,\Big]\,,
\end{split}
\end{equation}
\begin{equation}\label{Abeso7}
\begin{split}
A^{(\beta)}_{m\,ijkl} & \eql (u_1^2-v_1^2)\, \bar\eta_{[i}\Gamma_m{}^a\eta_j\bar\eta_k\Gamma^a\eta_{l]}\\[6 pt]
 &\quad +  (u_1u_2-v_1v_2)\,\Big[\coeff {4i}3\, \xi^a\,\bar\eta_{[i}\Gamma_m{}\eta_j\bar\eta_k\Gamma^a\eta_{l]}\\ &\hspace{1.5 in} -\coeff 2 9\,(3+\xi)\,\bar\eta_{[i}\Gamma_m{}^a\eta_j\bar\eta_k\Gamma^a\eta_{l]}+\coeff 2 9 \, {\xi_m\xi^a\over (3-\xi)}\,\bar\eta_{[i}\Gamma_m{}^a\eta_j\bar\eta_k\Gamma^a\eta_{l]}
\Big] \\ 
 & \quad + (u_2^2-v_2^2)\,\Big[ -\coeff {16i}{3} \,\xi^a\,\bar\eta_{[i}\Gamma_m\eta_j\bar\eta_k\Gamma^a\eta_{l]}
 -\coeff {2i} 9 \,(9+\xi) \,\xi_m\,\bar\eta_{[i}\Gamma^a\eta_j\bar\eta_k\Gamma^a\eta_{l]}\\ & \hspace{1.25in} +\coeff{2i} 3\, {\xi_m\xi^a\xi^b\over (3-\xi)} \,\bar\eta_{[i}\Gamma^a\eta_j\bar\eta_k\Gamma^b\eta_{l]}+\coeff 4 9 (15+2\xi)\,\bar\eta_{[i}\Gamma_m{}^a\eta_j\bar\eta_k\Gamma^a\eta_{l]}\\ & \hspace{1.5 in}-\coeff 2 9\, {\xi_m\xi^a\over(3-\xi)}\,\bar\eta_{[i}\Gamma^{ab}\eta_j\bar\eta_k\Gamma^b\eta_{l]}
 \Big] \,.
\end{split}
\end{equation}

The $U$-rotation is now effectively a substitution in \eqref{Aalso7} and \eqref{Abeso7} of the form
\begin{equation}\label{}
\begin{split}
\bar\eta_i\Gamma^a\eta_j \quad   \longmapsto \quad \cos(2\vartheta)\,\Gamma^a_{AB}-i\,\sin(2\vartheta)\,\bfs\xi^b\,\Gamma^{ab}_{AB}+2\sin^2\vartheta
\,\bfs\xi^a\,\bfs\xi^b\,\Gamma^b_{AB}\,,\end{split}
\end{equation}
and
\begin{equation}\label{}
\bar\eta_i\Gamma^{ab}\eta_j\quad   \longmapsto \quad \Gamma^{ab}_{AB}+2\, i \sin(2\vartheta)\,\bfs\xi^{[a}\Gamma^{b]}_{AB} +4\sin^2\vartheta \,\bfs\xi^c\bfs\xi^{[a}\,\Gamma^{b]c}_{AB}\,.
\end{equation}
Note that, in particular, we have
\begin{equation}\label{}
\xi^a\,\bar\eta_i\Gamma^a\eta_j \quad   \longmapsto \quad \xi^a\,\Gamma^a_{AB}\,,
\end{equation}
as expected from a rotation about $\xi^a$.

After tedious algebra, we find
\begin{equation}\label{rottens}
\begin{split}
{\cals A}_{m\,ABCD}(\alpha,\beta)\eql 
  \,P_1\,\bxi_m\,\Ga^a_{[AB}\Ga^a_{CD]} +P_2& \,\bxi^a\,\Ga_{m[AB}^{\phantom{a}}\Ga^a_{CD]}+P_3\,\bxi_m\bxi^a\bxi^b\,\Ga^a_{[AB}\Ga^b_{CD]}\\[6 pt]
  &+Q_1\,i\,\Ga_{m}{}^{a}_{[AB}\Ga^a_{CD]}+Q_2\,i\,\bxi_m\bxi^a\,\Ga^{ab}_{[AB}\Ga^b_{CD]}\,,
\end{split}
\end{equation}
which has the correct form \eqref{Ad11} of a solution to the {\it GVP} in $d\!=\!11$. 
Indeed, the first three terms on the right hand side, with 
\begin{equation}\label{}
\begin{split}
{P_1\over P_0}\eql \coeff 2 3\,H^2\,,\qquad {P_2\over P_0}\eql -2H\,,\qquad {P_3\over P_0}\eql { -} 2H(H-1)\,,
\end{split}
\end{equation}
and the common factor
\begin{equation}\label{}
P_0\eql -\coeff 3 4\,(\alpha+4\beta)\,{\tau(1+9\tau)\over (1+21\tau)^2}\,\sqrt{(21+\xi)(3-\xi)}\,,
\end{equation}
combine correctly to
\begin{equation}\label{}
-\coeff 3 4 (\alpha+4\beta)\,(S^{-1}\Do_m S)_{(ab)}\,\Ga^a_{[AB}\Ga^b_{CD]}\,,
\end{equation}
confirming the relation $\alpha + 4\beta =1$.
This can be checked by evaluating the background derivative of the 
siebenbein \eqref{so7pviel} using two identities
\begin{equation}\label{}
\begin{split}
\Do_m\xi & \eql 2\sqrt{(21+\xi)(3-\xi)}\,\bxi_m\,,\\[6 pt]
\Do_m\bxi^a & \eql \sqrt{{3-\xi\over 21+\xi}}\,(\eo_m{}^a-\bxi_m\bxi^a)\,,
\end{split}
\end{equation}
that follow  from the definition \eqref{defofxi}.

The coefficients in the remaining two terms in  \eqref{rottens} are considerably more complicated:
\begin{equation}\label{theQs}
Q_1\eql - \coeff1 8\,(\alpha-4\beta){1+21\tau\over 1-3\tau}\,{1\over H}+\coeff 1 8\,(\alpha+4\beta)\,{1-3\tau\over 1+21\tau}\,H\,,
\end{equation}
and
\begin{equation}\label{}
\begin{split}
Q_2\eql & \beta\,{1+21\tau\over 1-3\tau}\,{1\over H^2}+\coeff 1 8 (\alpha-4\beta){1+21\tau\over 1-3\tau}\,{1\over H}\\[6 pt]
& -\coeff 1 4 \,(\alpha+4\beta)\,{1+18\tau+225\tau^2\over(1-3\tau)(1+21\tau)}\\[6 pt]
& -\coeff 1 8 \,(\alpha+4\beta) \,{1-3\tau\over 1+21\tau}\,H+\coeff 1 4 (\alpha+4\beta)\,{1-3\tau\over 1+21\tau}\,H^2\,.
\end{split}
\end{equation}
Comparing \eqref{rottens} with \eqref{Ad11}, we read off $X_{m\,a}$ and confirm 
that $X_{m\,abc}=0$. By contracting the former with the inverse siebenbein, 
see \eqref{Xtwo}, and setting $\alpha$ and $\beta$ to the required values 
\eqref{specalpha}, we find that
\begin{equation}\label{}
X_{a|b}\eql X_0\,\delta^{ab}+X_2\,\bxi^a\bxi^b\,,
\end{equation}
where
\begin{equation}\label{}
\begin{split}
X_0 & \eql -\frac{(21 \tau +1)^2-7 H^2 (1-3 \tau )^2}{56 H^{4/3} (1-3 \tau )^{7/6} (21 \tau
   +1)^{5/6}}\,,\\[6 pt]
X_2  & \eql   \frac{\left(H^2-1\right) \left(2 H^2 (1-3 \tau )^2-(21 \tau +1)^2\right)}{8 H^{4/3}
   (1-3 \tau )^{7/6} (21 \tau +1)^{5/6}}\,. 
   \end{split}
\end{equation}
Decomposing $X_{a|b}$ into irreducible components \eqref{XXfour}, we get
\begin{align}\label{so7px0}
X& \eql X_0+\coeff 1 7 \,X_2\,,\\[6 pt]
X^{\,\Yboxdim4pt\yng(2)}_{ab} & \eql -\coeff 1 7 \,X_2\,\delta_{ab}+X_2\,\bxi_a\bxi_b\,.\label{so7px1}
\end{align}

From \eqref{projfl} we now obtain the flux along the entire $\rm SO(7)^+$ family,
\begin{equation}\label{thefluxso7p}
f\eql\frac{\sqrt{2}\, H^{2/3} \left(H^2 (1-3 \tau )^2-198 \tau ^2-36 \tau +2\right)}{(1-3
   \tau )^{7/6} (21 \tau +1)^{5/6}}\,,\qquad F_{abcd}\eql 0\,.
   \end{equation}
Once more, at the $\rm SO(8)$ point with $\tau=0$ and $H=1$, we reproduce \eqref{FrRsol}. 
At the $\rm SO(7)^+$ point, we have $99\tau^2 + 18\tau =1$, and therefore the expression
for $f$ simplifies to
\begin{equation}\label{}
f\eql 2^{1/2}\cdot 5^{3/4}\,\Delta^2\,,
\end{equation}
which agrees  with the known solution, see Table I in \cite{deWit:1984nz}.\footnote{The                         comparison involves setting the same overall normalization of the solutions, see Section~\ref{seccrpt} below.} As required by consistency, the Freund-Rubin parameter 
$f_0 = f \Delta^{-2}$ becomes $y$-independent at the critical point. Hence the standard inhomogeneous solution does reproduce the correct flux at the $\rm SO(7)^+$ point as well. 

However, since $X_2$ does not vanish outside the maximally supersymmetric point, we conclude that in this example {\it  the standard inhomogenous solution   does not have the correct tensor structure  \eqref{su8ABA}}. This can be seen even more directly by comparing the last two terms in \eqref{rottens} with the $e_{ma}f$ term in \eqref{su8ABA}. In order that the matrix
\begin{equation}\label{}
Q_1\eo_{ma}+Q_2\,\bxi_m\bxi_a\,,
\end{equation}
be {\it proportional\/} to $e_{ma}$, we must have 
\begin{equation}\label{}
{Q_2}- \left({1\over H}-1\right)\,Q_1\eql 0\,.
\end{equation}
Evaluating the left hand side we get
\begin{equation}\label{}
(\alpha+4\beta)\frac{\left(H^2-1\right) \left(2 H^2 (1-3 \tau )^2-(21 \tau
   +1)^2\right)}{8 H^2 (3 \tau -1) (21 \tau +1)}\eql 0\,,
\end{equation}
which can vanish at each point on $S^7$ when either $H\equiv 1$, which is the maximally supersymmetric solution, or when $\alpha+4\beta=0$. However, the latter cannot be satisfied given  \eqref{specalpha}. In the next section we will argue that this appears to be a 
generic feature of the $\cals A_{m\,ijkl}(\alpha,\beta)$ tensor in \eqref{theAzero}. 
To see that there is no discrepancy here at all,
we verify explicitly that the standard inhomogeneous solution 
can be shifted as in Section~\ref{secthree}, such that one obtains a solution satisfying 
the correct tensor structure condition. From the general solution to the homogeneous 
equations found in Sections~\ref{sectwo} and \ref{secthree}, and using the $\rm SO(7)^+$ invariance to limit the allowed $\alpha$-parameters, we find that the correction must 
be of the form
\begin{equation}\label{}
\delta\,\cals A_{m\,ijkl}\eql \, (\alpha_m{}^n+\coeff 3 7 \,\delta_m{}^n)\,\cals A^{(\alpha)}_{n\,ijkl}-
(\coeff 1 4\,\alpha_m{}^n+\coeff 3{28}\,\delta_m{}^n)\,\cals A^{(\beta)}_{n\,ijkl}\,.
\end{equation}
The result for the $U$-rotated tensor can be read-off from the $\alpha$ and $\beta$ 
components of $\cals A_{m\,ABCD}(\alpha,\beta)$ in \eqref{rottens}. We then find that 
the $S^{-1}\Do_mS$ terms cancel, as they should. Imposing the correct tensor 
structure condition yields a unique solution
\begin{equation}\label{}
\alpha_m{}^n (\xi,\tau) \eql a_0(\xi,\tau)\,\delta_m{}^n+a_2(\xi,\tau) \,\bxi_m\bxi^n\,,
\end{equation}
where
\begin{equation}\label{}
\begin{split}
a_0 & \eql -\coeff 1 7\,a_2\eql -\frac{\left(H^2-1\right) \left(2 H^2 (1-3 \tau )^2-(21 \tau +1)^2\right)}{14 (21
   \tau +1)^2}\,.
\end{split}
\end{equation}
Note that $\alpha_m{}^m=0$ as required, however, the correction breaks the $\rm SO(7)$ `background covariance' in the sense of the comment after \eqref{getalpgha}.

\section{Numerical examples}
\label{secseven}
\setcounter{equation}{0}

In this section, we summarize some numerical tests that led us to reexamine the proof of the consistent truncation in  \cite{deWit:1986iy}, and  discuss the standard inhomogenous solution and fluxes  for additional critical points.  We also found `numerical explorations' to be quite helpful  in developing   analytic arguments in  Sections \ref{sectwo}-\ref{secfive}. 
  The term `numerical' is used here in a wide sense; it means either an actual  numerical solution to the system of equations given by the consistent truncation    ans\"atze, or simply an explicit  evaluation (mixed numerical and analytic)  of expressions  using specific coordinates on the internal manifold and a particular representation of the internal $\Gamma$-matrices. 

It is important to note that the construction of the lift for the metric and the fluxes at a given point on the scalar manifold, $\rm E_{7(7)}/SU(8)$, is   `algebraic' with respect to the internal space. This is of course manifest for the metric, cf.\ \eqref{metrivfor}, but is also true for the fluxes,  the simple reason  being that the background covariant derivative always acts  on the Killing spinors or vectors on $S^7$  and,   by   virtue of \eqref{kspeqs} or  \eqref{derKillVect}, respectively, is effectively an algebraic operation. Hence, all equations given by the {ans\"atze} can be evaluated and then solved at each  point on $S^7$  independently, and the solution involves only algebraic operations.

\subsection{Preliminaries} 
In the following, we use  stereographic coordinates on $S^7$ and the Killing spinors that are obtained as follows:\footnote{For a systematic discussion, see \cite{PvanNreview:2008} and the references therein.} 
Represent $S^7$ as the surface, $(X^1)^2+\ldots+(X^8)^2= m_7^{-2}$, in $\RR^8$. The  stereographic coordinates, $y^a\in\mathbb{R}\; (a=1,\ldots,7)$, are defined  by
\begin{equation}\label{}
\begin{split}
X^i & \eql m_7^{-1}{2 y^i\over 1+(y^a)^2}\,,\qquad 
X^8   \eql m_7^{-1}{1-(y^a)^2\over 1+(y^a)^2}\,,\qquad i\eql 1,\ldots \,,7\,,
\end{split}
\end{equation}
and we use
\begin{equation}\label{}
\eo^a\eql -{2\,m_7^{-1}\,dy^a\over 1+(y^a)^2}\,,\qquad a=1,\ldots,7\,,
\end{equation}
as the background siebenbein. Then the    spin-connection 1-forms are ${\overset{_{\phantom{.}\circ}}{\omega}{}}{}^{ab}\eql -m_7\,(y^a\eo^b-y^b\eo^a)$, such that  the Ricci tensor is ${\overset{_{\phantom{.}\circ}}{R}{}}_{ab}\eql  6\,m_7^{2}\,\delta_{ab}$. 

The  matrix, $\eta=(\eta^I{}_A)$, of the Killing spinors is 
\begin{equation}\label{defeta}
\eta(y) \eql { \bfs 1+i\,y^a\,\Gamma^a\over\sqrt{1+(y^b)^2}}\,.
\end{equation}
It is easy to check that $\eta^I$'s satisfy the Killing spinor equation \eqref{kspeqs}, and that $\eta$ is a real orthogonal matrix. At  the North Pole, $y^a=0$, we have $\eta^I{}_A\eql\delta^I{}_A$
and ${\overset{_{\phantom{.}\circ}}{\omega}{}}{}^{ab}\eql 0$.

We use the $\rm SO(7)$ gamma matrices, $\Gamma^a_{AB}$, $a=1,\ldots,7$, that are antisymmetric and purely imaginary and satisfy 
\begin{equation}\label{}
\Gamma^a\Gamma^b+\Gamma^b\Gamma^a\eql 2\,\delta^{ab}\,\bfs 1\,,
\end{equation}
\begin{equation}\label{gamma7}
 \Gamma^7=i\,\Gamma^{123456}\,.
\end{equation}
If needed, explicit representations  with these properties can be found in Appendix C.1 of \cite{hep-th/9904017}, 
 or in Appendix C of \cite{LPTENS 79/6}. Note that the latter is for the negative Euclidean signature and gives the opposite sign in  \eqref{gamma7}.

\subsection{Initial numerical tests}

In our initial tests, we looked at the orginal flux formula \eqref{fluxeqs} for two non-supersymmetric critical points of the scalar potential of $\cN\!=\!8$, $d\!=\!4$ supergravity: the perturbatively unstable   $\rm SU(4)^-$ point and the perturbatively stable  $\rm SO(3)\times SO(3)$  point. The same calculation performed for a random scalar 56-bein yields similar results.

Starting with  \eqref{fluxeqs}, which is (7.5) in \cite{deWit:1986iy}, we evaluate the trace  over $m=q$.
Since the flux $F_{mnpq}$ should be totally antisymmetric, we then get
\begin{equation}\label{firstest}
\coeff{12}{7}(if)\,g_{np}\eql { -} i\coeff {\sqrt 2}{480}\,\Delta^4\,\epsilon_{pqrstuv}\,e^q_{ij}(e^r\bar e^se^t\bar e^ue^v)_{kl}\cals A_n{}^{ijkl}(\alpha,\beta)\,.
\end{equation}
The left hand side is now proportional to the metric tensor, and thus its contraction with the inverse metric tensor $\Delta ^{-1}g^{mn}$ in \eqref{metrivfor} should yield a result proportional to the identity matrix.
To test that, we would fix a point on $S^7$, either at  the North Pole or at some random value  of the stereographic coordinates, and evaluate numerically:\vspace{-4 pt}
\begin{itemize}
\item [(i)]  the inverse tensor in \eqref{metrivfor}, and
\vspace{-6 pt}
\item [(ii)] the tensor defined by the right hand side in \eqref{firstest}  for arbitrary values of $\alpha$ and $\beta$. 
\end{itemize}
\vspace{-4 pt}
The result is that  for the $\rm SU(4)^-$ point, the contraction between the tensors (i) and (ii) is not proportional to the unit matrix, while for the $\rm SO(3)\times SO(3)$ point, the tensor in (ii) is not even symmetric.
In both examples, the undesired terms are proportional to $\alpha+4\beta$ and do {\it not\/} vanish for the standard inhomogenous solution \eqref{specalpha}. In retrospect, they arise because $\cAo_{m\,ijkl}$ does not satisfy the correct tensor structure condition \eqref{su8ABA}.

\subsection{Tensor structure tests}
\label{sectenstr}

The main purpose of a more systematic numerical exploration is to determine the structure of the $\cAo_{m\,ABCD}$ tensor obtained by the $U$-rotation of the $\cAo_{m\,ijkl}$ tensor of the standard inhomogenous solution. 

Once more we take a  scalar 56-bein, $\cals V$, for one of the  critical points and choose a random point on $S^7$. We then evaluate numerically the metric tensor, $g_{mn}$, and the warp factor, $\Delta$, using \eqref{metrivfor}. By taking the matrix square root of the metric tensor, we find the metric vielbein, $e_m{}^a$, and its inverse.  Then  \eqref{aligngv} 
becomes a quadratic equation for the $\rm SU(8)$-matrix, $U$, with the latter   determined up to an overall sign that cancels in  \eqref{Utransform}. The resulting $\cAo_{m\,ABCD}$  tensor is expanded in the canonical basis, cf.\ \eqref{Ad11}, and we  read-off the expansion coefficients, $X_{m\,a}$ and $X_{m\,abc}$.  Finally we evaluate $X_{a|b}$ and $X_{a|bcd}$ in \eqref{Xtwo} and decompose them into irreducible $\rm SO(7)$ components \eqref{XXfour}.

\begin{table}[t]
\begin{center}
\scalebox{0.8}{
\begin{tabular}{@{\extracolsep{25 pt}}c c c c c c c c}
\toprule
\noalign{\smallskip}
\quad$\cal V$ & Symmetry & $X$ & $X_{abcd}^{\,\Yboxdim4pt\yng(1,1,1,1)}$  & $X^{\,\Yboxdim4pt\yng(2)}_{ab}$ & $X^{\,\Yboxdim4pt\yng(1,1)}_{ab}$ & $X_{abcd}^{\,\Yboxdim4pt\yng(2,1,1)}$ & $\delta^{\phantom{{\,\Yboxdim4pt\yng(1,1)}}}_{a[b}\tilde X_{cd]}^{\,\Yboxdim4pt\yng(1,1)}$ \\
\noalign{\smallskip}
\midrule
\noalign{\smallskip}
S0600000 & $\rm SO(8)$ & * \\
S0668740 & $\rm SO(7)^-$ & * & * \\
S0698771 & $\rm SO(7)^+$ & * &   & * \\
S0719157 & $\rm G_2$ & * & * & * & * & * & * \\
S0779422 & $\rm SU(3)\times U(1)$ & * & * & * & * & * & * \\
S0800000 & $\rm SU(4)^-$ & * & * & * \\
S0880733 & $\rm SO(3)\times SO(3)$& * & * & * & * & * & * \\
S1200000 & $\rm U(1)\times U(1)$  & * & * & * & * & * & * \\
S1400000 & $\rm SO(3)\times SO(3)$ & * & * & * & * & * & * \\
\bottomrule
\end{tabular}
}
\caption{\label{numcheks} 
{ Irreducible components in ${\mathcal{A}}{}_{m\,ABCD}\big(\coeff 4 7,\coeff 3 {28}\big)$ at some critical points
}}
\end{center}
\end{table}

Our results are summarized in Table \ref{numcheks}, where the star indicates that a given irreducible component does not vanish.  The critical points are listed in the first column using the labelling scheme in \cite{Fischbacher:2011jx} that is based on the value of the cosmological constant. The second column gives the symmetry of each point, which is perhaps more recognizable than the label. The reader may consult  \cite{Fischbacher:2011jx} for additional information about and references for each point. The last four columns are the components 
that violate the tensor structure condition: we see that already the highly symmetric
$\rm G_2$ solution gives rise to all possible tensor structures.
This table includes all `old' critical points and two `new' ones, 
{\rm \small S0880733} and {\small S1200000}, first found numerically in \cite{Fischbacher:2009cj,Fischbacher:2011jx} and then further investigated in  \cite{Fischbacher:2010ec}.

\subsection{Critical points}
\label{seccrpt}

The calculation in Section~\ref{sectenstr} also gives us the flux, $f$, which can be readily compared with the known lifts of critical points. Recall that at
 a critical point, the flux, $f$, along $AdS_4$ is by conformal invariance and the Bianchi identity of the form $f=f_0\Delta^{2}$, where $f_0$ is a constant. We determine numerically the value of $f_0$ from the coefficient $X_{ma}$ in $\cAo_{m\,ABCD}$ using \eqref{fformula}. Then we verify that $f_0$ is indeed constant by performing the same calculation for two or more points on the sphere. The results are listed in Table~\ref{f0flux}.
 
Next we compare our numerical results with the known solutions. This requires ensuring that a solution \eqref{themetr}-\eqref{fluxsol} obtained from the lift and a solution we compare it 
with have the same overall normalization. The potential mismatch between the normalizations comes from the fact that  the field equations of $d\!=\!11$ supergravity are invariant 
under the rescaling
\begin{equation}
g_{MN}\rightarrow \lambda\, g_{MN}\,, \qquad
F_{MNPQ}\rightarrow \lambda^{3/2}F_{MNPQ} \,,
\end{equation}
where $\lambda $ is a constant. This rescaling preserves the form of a solution \eqref{themetr}-\eqref{fluxsol}, but changes the radius of $AdS_4$ and the overall scale of the internal metric and the flux. Let us also note that the relative normalization between the $AdS_4$ and the internal parts of the metric and the flux is completely fixed by the equations of motion.  In particular, from a linear combination of the Einstein equations, we have\footnote{See, e.g.\ (3.7) in 
   \cite{deWit:1984nz}. As shown there, this equation implies the relation 
   $15m_4^2 \gamma^{-1/2} - f_0^2 \gamma^{-5/3} = 42 m_7^2$, explaining the 
   powers of $\gamma$ appearing in \eqref{tblmf}.}
\begin{equation}\label{foform}
R_m{}^m+\coeff 5 4\,R_\mu{}^\mu\eql f_0^2\Delta^4\,,
\end{equation}
which effectively is the equation of motion we are testing here.

\begin{table}[t]
\begin{center}
\scalebox{0.8}{
\begin{tabular}{@{\extracolsep{25 pt}}c c c c c}
\toprule
\noalign{\smallskip}
\quad$\cals V$ & Symmetry & $-{\cals P}_*/g^2$ & $f_0/m_7$  & Refs  \\
\noalign{\smallskip}
\midrule
\noalign{\smallskip}
S0600000 & $\rm SO(8)$ &  6 & 4.242641 & \cite{EFI 80/35-CHICAGO} \\
S0668740 & $\rm SO(7)^-$ & 6.68740 & 4.941059 & \cite{CERN-TH-3394,deWit:1983gs}\\
S0698771 & $\rm SO(7)^+$ & 6.98771 & 4.728708 & \cite{deWit:1984va,deWit:1984nz}\\
S0719157 & $\rm G_2$ & 7.19157 & 5.085212  & \cite{deWit:1984nz} \\
S0779422 & $\rm SU(3)\times U(1)$ & 7.79422 &  5.511352 & \cite{Corrado:2001nv}  \\
S0800000 & $\rm SU(4)^-$ & 8 &  5.656854 & \cite{Pope:1984bd}\\
S0880733 & $\rm SO(3)\times SO(3)$& 8.80733 & 6.227729 & $-$\\
S1200000 & $\rm U(1)\times U(1)$ & 12 & 8.485281  & $-$\\ 
S1400000 & $\rm SO(3)\times SO(3)$& 14 &  9.899495 & \cite{BKPWtoappear} \\
\bottomrule
\end{tabular}
}
\caption{\label{f0flux} 
{The $f_0$-flux at some critical points.  }}
\end{center}
\end{table}

The solution obtained from the lift comes with a particular normalization determined by the explicit embedding of the $d\!=\!4$ solution in  eleven dimensions.  Specifically,  the radius, $L$, of  $AdS_4$ in \eqref{themetr}  is given by 
\begin{equation}\label{radeqs}
L^2\equiv m_4^{-2}= -{3\over \cals P_\star}\,,
\end{equation}
where  $\cals P_*$ is  the  value of the scalar potential of the $\cN\!=\!8$ theory \cite{de Wit:1982ig},
\begin{equation}\label{scalpot}
\cP\eql -g^2\big(\coeff{3}{4}\left|A_1{}^{ij}\right|^2-\coeff{1}{24}\left|A_{2i}{}^{jkl}\right|^2\big)\,,
\end{equation}
at the critical point. The normalization of the internal metric and the flux are in turn determined by  the lift formulae \eqref{metrivfor} and \eqref{projfl}.  

To test  the first four points, we use solutions summarized in Table I 
in \cite{deWit:1984nz}, where we find
 \begin{equation}\label{tblmf}
m_4^2\eql a\,m_7^2\,\gamma^{1/2}\,,\qquad f_0\eql b\,m_7\,\gamma^{5/6}\,.
\end{equation}
The values of the constants $a$ and $b$  depend on the critical point under 
consideration and can be read off from Table I in \cite{deWit:1984nz}, while $\gamma$ 
is an arbitrary parameter that sets the overall normalization of the solution. 
For each critical point we find the correct $\gamma$ by solving \eqref{radeqs} with $m_4$ in \eqref{tblmf}. Then we use this particular value of $\gamma$ to evaluate $f_0$ in  \eqref{tblmf} and in all four cases find a complete agreement with the numerical values in Table \ref{f0flux}.

The $\rm SU(3)\times U(1)$ solution is given in \cite{Corrado:2001nv}.  From 
(4.29) and (4.33) in that paper we  get\footnote{\label{foot}There is a difference in the normalization of the flux in \cite{Corrado:2001nv} and in this paper,  $F_{(4)} =\sqrt 2\,F_{(4)}^{\rm CPW}$.} 
\begin{equation}\label{su3f0}
f_0\eql \frac{3^{7/4}}{2^{3/2} }{1\over L}\,.
\end{equation}
At the critical point,
\begin{equation}\label{}
\cals P_*\eql -{3^{5/2}\over 2}\,g^2\,,
\end{equation}
so from \eqref{radeqs}, and recalling that $g=\sqrt 2\,m_7$, we get
\begin{equation}\label{}
{1\over L}\eql 3^{3/4} \,m_7\,.
\end{equation}
Substituting this in \eqref{su3f0}  yields
\begin{equation}\label{}
f_0\eql {3^{5/2}\over 2^{3/2}}\,m_7\approx 5.51135\, m_7\,,
\end{equation}
which agrees with the numerical value in Table \ref{f0flux}. 

The solution at the $\rm SU(4)^-$ critical point was found in \cite{Pope:1984bd}. For the comparison we use the explicit formulae (4.70) and (4.71) in \cite{Bobev:2010ib}, which after rescaling the flux (see, footnote \ref{foot}) read
\begin{equation}\label{PWsol}
ds_{11}^2\eql ds^2_{AdS_4}+\ldots\,,\qquad F_{(4)}\eql \sqrt{{3\over 2}}{1\over L}\,{\rm vol}_{AdS_4}+\ldots\,.
\end{equation}
However, the metric obtained from the lift \eqref{metrivfor} has a nonvanishing constant warp factor, $\Delta=2^{-2/3}$. Reintroducing this warp factor in \eqref{PWsol} by rescaling the metric by $\Delta^{-1}$ and the flux by $\Delta^{-3/2}$, we get $f_0=\sqrt 6/L$. Then, using $\cP_*=-8g^2$, and normalizing the $AdS_4$ radius according to  \eqref{radeqs}, we get
\begin{equation}\label{}
f_0\eql 4\sqrt 2\,m_7\approx 5.65685\, m_7\,,
\end{equation}
which is the same as the numerical value obtained from the lift.

Finally, the flux, $f_0$, for the $\rm SO(3)\times SO(3)$ critical point has been calculated 
in \cite{BKPWtoappear} by solving \eqref{foform}, where the metric is obtained from the 
lift formula \eqref{metrivfor}. Once more we find that it agrees with the numerical result 
in Table \ref{f0flux}. Analytic solutions for the remaining two points in Table \ref{f0flux} 
are not known explicitly in closed form.

In Table~\ref{f0flux} we have also listed the values of the scalar potential at each of the critical points. We see that  there is a universal relation between $\cals P_*$, or equivalently the radius of $AdS_4$,  and $f_0$, which in our normalization reads
\begin{equation}\label{Ptof0}
-{\cals P_*\over g^2}\eql \sqrt 2\,{ f_0\over m_7} \,.
\end{equation}
In principle, this relation is a consequence of \eqref{foform}, but we are not aware of any simple proof of it. The difficulty here is that the components of the Ricci tensor in \eqref{foform} are for the full $d\!=\!11$ metric.

Curiously, the relation \eqref{Ptof0} holds for some other field configurations, for instance in the entire $\rm SO(7)^-$ invariant sector, see \eqref{Pinso7m}, \eqref{Delso7m} and \eqref{ffromlift}. However, it is not valid in general, in particular, at a generic point in the $\rm SO(7)^+$ sector, where $f\Delta^{-2}$, c.f.\ \eqref{so7pwarp} and  \eqref{thefluxso7p}, has a nontrivial dependence on the sphere coordinates which cancels out only at the two critical points.

While our tests in this section were limited only to solutions corresponding to the critical points, and we looked only at the flux component along $AdS_4$, it is clear that the agreement we have found is a striking confirmation of the lift formulae for the flux.

\section{Conclusions and outlook}
\label{seceight}
\setcounter{equation}{0}

In this paper we have clarified the structure of the equations given in \cite{deWit:1986iy}
(that is, the {\it GVP} and the $\frak A$-equations) which characterize consistent truncations 
of eleven-dimensional supergravity on $AdS_4\times S^7$ to gauged supergravity. 
We have revealed a hidden degeneracy in these equations and demonstrated that this degeneracy is precisely what is needed in order to remove apparent discrepancies 
arising in the comparison between the $d\!=\!4$ and $d\!=\!11$ expressions, and to 
recover the correct tensor structure of the fluxes required by the $d\!=\!11$ theory for 
{\em any} given non-trivial solution of the $d\!=\!4$ theory. Furthermore, we have clarified 
the status of the non-linear ans\"atze for the fluxes, and shown that these constitute 
invariants of the consistency equations. 

These `flux lift formulae' can now be put to practical 
use, and we have presented several non-trivial tests, both analytic and numerical. It is also clear from our discussion that on the one side an  analytic calculation of the fluxes based on those formulae is quite difficult and cumbersome, though perhaps it can be simplified in particular examples  by a judicious choice of coordinates and the  Killing vectors/spinors.  On the other side, a numerical calculation is reasonably straightforward and very likely may be sufficient to determine properties of the full solutions one might be interested in.

Let us also remark  that the degeneracy problem discussed in this paper does not arise 
for the `mixed'  flux components: $F_{\mu bcd}$, $F_{\mu\nu bd}$, etc., which will no 
longer vanish for $x$-dependent solutions of the $d\!=\!4$ theory; these can therefore be determined unambiguously from the corresponding
formulas given in \cite{deWit:1985iy} and \cite{deWit:1986iy,deWit:1986mz}. 

Finally, our results may also be relevant in the context of the $AdS_5\times S^5$
compactification of IIB supergravity, for which the analog of the metric lift formula
\eqref{metrivfor} is known, but a complete proof of the consistency is still lacking.
{\em Mutatis mutandis} we anticipate that the techniques developed here
on the basis of \cite{deWit:1986iy} will also apply to this case.

\bigskip
\bigskip
\leftline{\bf Acknowledgements}
\smallskip
We would like to thank Nikolay Bobev, Bernard de Wit and Nick Warner for discussions. 
The work of KP is supported in part by DOE grant DE-FG03-84ER-40168.  HN would like to thank the Simons Center for Geometry and Physics at Stony Brook University for hospitality in May 2011 when this project was begun, while KP is grateful
to the Albert Einstein Institute in Potsdam for hospitality during two visits at the later stages of the project.



\end{document}